\documentclass[twocolumn]{openjournal}
\def\draftversion{0} 
\usepackage{amssymb,latexsym,graphicx,natbib,eufrak,times,amsmath,xspace,ifthen, xcolor}
\usepackage[normalem]{ulem}
\usepackage{orcidlink}

\usepackage{hyperref}
\definecolor{linkblue}{rgb}{0,0.4,0.6}
\hypersetup{colorlinks, linkcolor={linkblue}, citecolor={linkblue}, urlcolor={linkblue}}

\ifthenelse{ \draftversion > 0 }{
  \newcommand{\so}[1]{\color[rgb]{0.6,0,0.6}\sout{#1}\color{black}} 

} {
  
  \newcommand{\so}[1]{}
}

\ifthenelse{\equal{\draftversion}{1}}{
  \newcommand{\sep}[1]{\par\begin{color}[rgb]{0,0.4,0}\begin{center}\hrule\end{center}\end{color}\par} 
  \newcommand{\todo}[1]{\begin{color}{red}\ \ifthenelse{\equal{#1}{}} {$\bullet\bullet\bullet$} {$\bullet$\ #1 $\bullet$}\end{color}} 
  \newcommand{\idea}[1]{\begin{color}[rgb]{0,0.4,0}\textit{#1}\end{color}} 
  \newcommand{\sk}[1]{\begin{color}[rgb]{0.6,0,0.6}#1\end{color}} 
  \newcommand{\toc}{\par\begin{color}[rgb]{0.6,0,0.6}\begin{center}\hrule\vspace{0.5mm}\begingroup\small\let\cleardoublepage\relax\let\clearpage\relax\mytoc\endgroup\vspace{0.5mm}\hrule\end{center}\end{color}\par} 

  }{
  \newsavebox{\trashcan}

  \newcommand{\sep}[1]{}
  \newcommand{\todo}[1]{}
  \newcommand{\idea}[1]{}
  \newcommand{\sk}[1]{}
  \newcommand{\toc}{}

  }
 \makeatletter\newcommand\mytoc{\@starttoc{toc}}\makeatother 
\long\def\symbolfootnote[#1]#2{\begingroup%
\def\thefootnote{\fnsymbol{footnote}}\footnote[#1]{#2}\endgroup} 

\newcommand{\eqn}[2][]{Equation#1~\ref{eqn:#2}} 
\newcommand{\fig}[2][]{Figure#1~\ref{fig:#2}}

\newcommand{\sect}[2][]{Section#1~\ref{sec:#2}}
\newcommand{\app}[2][]{Appendix#1~\ref{sec:#2}}

\newcommand{\bb}[1]{\ifmmode \mbox{\boldmath $ #1$} \else  \mbox{\boldmath $#1$} \fi}

\newcommand{\mh}{\ensuremath{\textrm{\,--\,}}}    
\newcommand{\dd}{\ensuremath{\,\mathrm{d}}}       
\newcommand{\U}[1]{\ensuremath{\mathrm{~#1}}}     
\newcommand{\e}[1]{\ensuremath{\times 10^{#1}}}   

\newcommand{\yr}{\U{yr}}
\newcommand{\Myr}{\U{Myr}}          \newcommand{\myr}{\Myr}
          
\newcommand{\pc}{\U{pc}}
\newcommand{\kpc}{\U{kpc}}
          
\newcommand{\Msun}{\U{M}_{\odot}}   \newcommand{\msun}{\Msun}

\newcommand{\cc}{\U{cm^{-3}}}
\newcommand{\K}{\U{K}}

\newcommand{\fgas}{\ensuremath{f_{\rm gas}}\xspace}      
\newcommand{\tdep}{\ensuremath{\tau_\mathrm{dep}}\xspace}        
\newcommand{\dms}{\ensuremath{\Delta\mathrm{SFR}_\mathrm{MS}}\xspace}        

\newcommand{\mstar}{\ensuremath{M_\star}\xspace}       

\newcommand{\ramses}{{\sc Ramses}\xspace}

\newcommand{\nh}{{\sc NewHorizon}\xspace}

\newcommand{\sxb}{Observatoire Astronomique de Strasbourg, Universit\'e de Strasbourg, CNRS UMR 7550, 11 rue de l'Universit\'e, F-67000 Strasbourg, France}
\newcommand{\usias}{University of Strasbourg Institute for Advanced Study, 5 all\'ee du G\'en\'eral Rouvillois, F-67083 Strasbourg, France}
\newcommand{\cea}{Laboratoire AIM Paris-Saclay, CEA/IRFU/SAp, Universit\'e Paris Diderot, F-91191 Gif-sur-Yvette Cedex, France}
\newcommand{\iap}{Institut d'Astrophysique de Paris, CNRS and Sorbonne Universit\'e, UMR 7095, 98 bis Boulevard Arago, F-75014 Paris, France}

\newcommand{\lam}{Aix Marseille Universit\'e, CNRS, LAM (Laboratoire d'Astrophysique de Marseille), 13388 Marseille, France}

\newcommand{\nottingham}{School of Physics and Astronomy, University of Nottingham, University Park, Nottingham NG7 2RD, UK}

\newcommand{\ilance}{ILANCE, CNRS – University of Tokyo International Research Laboratory, Kashiwa, Chiba 277-8582, Japan}
\newcommand{\kavlitokyo}{Kavli IPMU (WPI), UTIAS, The University of Tokyo, Kashiwa, Chiba 277-8583, Japan}

\begin{document}


\title[Starbursts hiding in the main sequence]{Starbursts hiding in the main sequence: a pathway toward quenching?}
\newcommand{\au}[3]{\author{#1\,\orcidlink{#2}\altaffilmark{#3}}}

\au{Florent~Renaud}{0000-0001-5073-2267}{1,2}
\au{Katarina~Kraljic}{0000-0001-6180-0245}{1}
\au{Jonathan~Freundlich}{0000-0002-5245-7796}{1}
\au{Benjamin~Magnelli}{0000-0002-6777-6490}{3}
\au{Matthieu~B\'ethermin}{0000-0002-3915-2015}{1}
\au{C\'edric~Accard}{0009-0005-9982-7239}{1}
\au{Diana~Ismail}{0009-0007-2281-4944}{1}
\au{Emanuele~Daddi}{0000-0002-3331-9590}{3}
\au{David~Elbaz}{0000-0002-7631-647X}{3}
\au{Laure~Ciesla}{0000-0003-0541-2891}{4}
\au{Garreth~Martin}{0000-0003-2939-8668}{5}
\au{Yohan~Dubois}{0000-0003-0225-6387}{6}
\au{S\'ebastien~Peirani}{0000-0001-6902-2898}{7,8,6}

\altaffiltext{1}{\sxb}
\altaffiltext{2}{\usias}
\altaffiltext{3}{\cea}
\altaffiltext{4}{\lam}
\altaffiltext{5}{\nottingham}
\altaffiltext{6}{\iap}
\altaffiltext{7}{\ilance}
\altaffiltext{8}{\kavlitokyo}



\begin{abstract}
Star-forming galaxies spend most of their lifetimes on the star-forming main sequence, which establishes a tight empirical and statistical relation between stellar mass and star-formation rate. Occasional episodes of rapid star formation can push them temporarily above this sequence, turning them into starbursts. Yet some galaxies display starburst-like traits --rapid, dense, and compact star formation-- while still remaining within the scatter of the main sequence. These ``starbursts in the main sequence'' (SBMSs) reveal the complexity and diversity of star formation modes, making them crucial for understanding how galaxies evolve and transition between different regimes. In this paper, we identify SBMSs in the cosmological simulation \nh and follow their evolution across time to uncover their physical origins and the role of this special regime in shaping galaxy evolution. We explain the existence of SBMSs by a comparatively earlier assembly of their stellar mass, driven in particular by more frequent and repeated mergers as the other galaxies, as well as exceptionnally productive starburst events triggered by these interactions. As a result, this regime appears preferentially --though not exclusively-- in the most massive galaxies. The SBMS behavior is not continuous within individual galaxies but instead arises intermittently as a short-lived ($\sim 30 \Myr$) evolutionary mode. Nevertheless, such SBMS episodes exist throughout cosmic time across the galaxy population, rooted in the inherently stochastic nature of galactic star formation histories. Contrary to common interpretations in the literature, the SBMS phase marks a transition between starburst and more quiescent star formation in only a small fraction of cases ($<25\%$). Compaction events do raise the star formation rate, but do not reduce the gas depletion time, i.e., the timescale for star formation. We find no evidence linking the SBMS regime to quenching via compaction (i.e., the blue-then-red-nugget pathway). The importance of the SBMS regime and the variety of evolutionary tracks into and out of it challenges attempts to summarize the evolution of star-forming galaxies with a single, universal scenario.
\end{abstract}


\section{Introduction}

The star formation rate (SFR) of galaxies has been empirically found to correlate with their stellar masses, along a relation called the main sequence of galaxy formation \citep{Noeske2007, Elbaz2007, Daddi2007, Rinaldi2025}. Although this relation evolves with redshift \citep{Whitaker2014, Speagle2014, Simmonds2025, Song2026}, it remains remarkably tight across cosmic time and stellar mass. The rare outliers with an excess of SFR for their mass (i.e., above the main sequence) are commonly referred to as starbursts, denoting an abnormal and likely short-lived activity, often ascribed to galaxy mergers (\citealt{Rodighiero2011, Elbaz2011, Schreiber2015, Clarke2025}, but see \citealt{Renaud2022} on mergers not triggering starbursts, and \citealt{Liu2025} on nuclear starbursts not triggered by mergers). Despite their high productivity, the rarity of starbursts indicates that most stars do not form in this regime \citep{Rodighiero2011}. Conversely, galaxies below the main sequence are often interpreted as being on a pathway toward quenching with a possibility of later rejuvenation \citep{Mancini2019}.

Statistical analyses of large surveys of galaxy populations, probed notably in the submillimetric and millimetric domains with ALMA, have revealed the existence of a hybrid category of galaxies with gas depletion times as short as those of starbursts, but lying within the scatter of the main sequence. These ``starbursts in the main sequence'' (SBMSs) exhibit compact dust and radio emission, elevated dust temperatures, and extreme far-infrared surface brightness \citep{Elbaz2018, Jimenez2019, Puglisi2019, Puglisi2021}, especially at the high-mass end \citep{Gomez2022}. All these aspects suggest a classification as starbursts, which is apparently incompatible with their belonging to the main sequence. The discovery of this population unveils a greater diversity (and likely complexity) in the regimes of star formation and the underlying physics of the main sequence. Among other questions, it raises the problem of transitions between star formation regimes, and whether such evolutions are quasi-adiabatic and smooth, or require a sudden shift in environmental conditions.

\citet{Magnelli2023} used MIRI/JWST data up to cosmic noon ($z\approx 2$) to find that a significant fraction of the main sequence galaxies displays a compact, dust-obscured starburst-like star forming region, but embedded in a larger stellar component. \citet{Lyu2025} further classified these galaxies as ``extended-compact'', representing $15\%$ of their sample but being twice as common among massive galaxies. They postulate that these galaxies could be the progenitors of ``compact-compact'' systems, for which both the star-forming and old stellar components appear compact, and perhaps on their way to becoming quiescent after a merger-triggered starburst phase \citep{Puglisi2019, Tarrasse2025}.

The enhancement of star formation activity during major mergers is not confined to the nucleus but rather spans large, kpc-scale volumes across the galaxy progenitors, including at high redshift (when resolved), as found observationally \citep[e.g.,][]{Schweizer2005, Whitmore2010, Mineo2014, Cortijo2017, Pan2019, Tsuge2020, Xu2021} and well reproduced in simulations \citep[e.g.,][]{dimatteo2008, Karl2010, Teyssier2010, Hopkins2013b, Renaud2014b, Renaud2019, Moreno2021, Li2022}. However, the central concentration of dense gas --particularly during the last phases of interactions-- leads to peaks of luminosity in the galactic nuclei \citep{Mihos1994, Ellison2025}. This process is invoked to describe the first steps of the compaction mechanism in post-merger, coalesced, systems, i.e. the central concentration of star-forming material \citep{Dekel2014, Puglisi2019, Lapiner2023}, possibly in connection with the SBMS regime. Yet, \citet{Magnelli2023} noted that the diversity of galaxy populations could imply a variety of evolutionary pathways, which must be understood to decipher the histories of galaxies in and out of the main sequence \citep[see also][]{Arango2025, Tarrasse2025}.

Toward this goal, \citet{Ciesla2023} proposed a quantification of the joint recent evolution of the SFR and the stellar mass. They employed spectral energy distribution (SED) fitting techniques to reconstruct the star formation history (SFH) of SBMSs identified as main sequence galaxies with dust-obscured, starburst-like activity in the GOODS-ALMA sample. They found no peculiarities in either the SEDs or the SFHs compared to other galaxies. Although SED fitting offers a unique access to the temporal dimension from observational data, parametric versions of this technique introduce artefacts by imposing the shape of the fitting functions \citep{Ciesla2017, Iyer2017, Lower2020}. Even when describing the SFH piecewise (i.e., with a non-parametric or, more precisely, a non-analytical approach like in \citealt{Ciesla2023}; see also \citealt{Ocvirk2006, Leja2019, Tacchella2020}), SED fitting suffers from uncertainties in stellar evolution and dust models, as well as degeneracies in the solutions \citep{Conroy2013, Caplar2019, Carvajal2025}.

In parallel, cosmological simulations are prime tools for understanding time evolutions and probing the diversity of histories among galaxies. \citet{Tacchella2016} used the {\sc Vela} cosmological zoom-in simulations to confirm the process of compaction through the ignition of a central starburst (``blue nugget''), followed by the formation of a stellar core and quenching of star formation activity. The concentration of gas is caused by loss of angular momentum via gravitational torques from mergers, accretion from large scales, or the formation of massive clumps in disk instabilities \citep{Lapiner2023}. In the \nh simulation, \citet{Martin2021} found that not all interaction-triggered enhancements of star formation would push the galaxies off the main sequence, especially at high reshift. However, the connection with the SBMS regime has not been explicitly studied, and hypotheses regarding the role of this regime in galaxy evolution remain highly speculative. Furthermore, although the high resolution of zoom-in simulations is essential to capture the physical conditions of star formation, it imposes limits on the modeled volume, and thus on the diversity of galaxies probed \citep[as underlined in][]{Zhang2025}.

In this paper, we use the cosmological simulation \nh \citep{Dubois2021} to propose a theoretical explanation for the nature and evolution of SBMS galaxies. This simulation combines a large volume, comprising a relatively wide diversity of galaxies, with the high resolution ($\approx 34 \pc$) needed to (partially) capture the conditions of star formation within these galaxies. The starting point of our study is a necessary clarification of the terms and definitions used (\sect{semantics}). We then present a series of diagnostics relying on population statistics as well as individual evolutions, to illustrate the nature of SBMS galaxies (\sect{stats}) and their possible evolution toward this peculiar regime (\sect[s]{origins} and \ref{sec:episodes}). After discussing the limitations of our approach (\sect{discussion}), we summarize our theoretical picture of this phenomenon (\sect{conclusions}).

\section{Method}

\subsection{Simulation}

We use the large-scale zoom-in cosmological simulation \nh\footnote{\href{http://new.horizon-simulation.org}{http://new.horizon-simulation.org}}
 presented in \citet{Dubois2021}, and the selection of star-forming galaxies from \citet{Kraljic2024}. We refer the reader to these papers for details, as only a summary is presented here. \nh is run with the \ramses code \citep{Teyssier2002}, with mass resolutions of $1.2\e{6} \msun$ and $1.3\e{4} \msun$ for the dark matter and stellar particles, respectively. Hydrodynamics is solved using the adaptive mesh refinement technique down to a resolution of $34 \pc$ in the densest regions, with heating from an ultraviolet background \citep{Haardt1996}, self-shielding of optically thick regions \citep{Rosdahl2012}, cooling from collisional ionization, excitation, recombination, bremsstrahlung, and Compton effect down to $10^4 \K$. Additional metal-line cooling allows the gas to cool further, down to $0.1 \K$ \citep{Dalgarno1972, Sutherland1993}.

Star formation is modeled in gas denser than $10 \cc$, with a non-uniform star formation efficiency $\epsilon_{\star}$ per free-fall time \citep{Krumholz2005, Padoan2011, Hennebelle2011, Federrath2012}. In short, this model assumes that the unresolved distribution of gas density within a given cell follows a log-normal distribution scaled by the local (resolved) turbulence. The Mach number and virial parameter measured at the scale of the cell then determine which density range of this unresolved distribution should be star forming, using best-fit relations and empirical calibrations from closed-box simulations of the interstellar medium (ISM). This range is then converted into a star formation efficiency for the individual cells, which, once multiplied by an arbitrary scaling factor and divided by the local free-fall time, finally gives the net efficiency $\epsilon_{\star}$ \citep[see][for details]{Dubois2021, Kraljic2024}. We discuss the implications of this model in \sect{subgrid}. 

Feedback from type-II supernovae is modeled by the injection of radial momentum following \citet{Kimm2014}. The simulation also includes the growth and dynamical evolution of massive black holes and their spin \citep{Dubois2014b}, and the associated active galactic nucleus (AGN) feedback via the injection of mass, momentum, and energy at low accretion rates (``radio mode''; \citealt{Dubois2010}), and of thermal energy only at higher rates (``quasar mode''; \citealt{Teyssier2011}).

Throughout the simulation, galaxies are identified using the clump finder {\sc AdaptaHOP} \citep{Aubert2004} applied to the stellar component. This yields a sample of approximately $150 \mh 1400$ star-forming galaxies (depending on redshift, with a peak at $z \approx 4$), with stellar masses between $5.7\e{5} \msun$ and $2.5\e{11} \msun$. Thus, our sample does not include analogs of the recently discovered progenitors of extremely massive galaxies undergoing early and rapid evolution \citep[see, e.g.,][]{Casey2024, Carniani2025}. This absence can be explained by either a too small simulation volume not covering the envrionmental conditions needed to host the formation of such galaxies, and/or missing physical processes, yet to be identified. On each snapshot, our working sample includes all the star-forming galaxies as in previous analyses of \nh in \citet{Dubois2021} and \citet{Kraljic2024}. By tracking the galaxies throughout the snapshots and monitoring the transfer of stellar mass between galaxies as in \citet{Tweed2009}, galaxy merger events are identified and catalogued \citep{Martin2021}.

\subsection{Measurements and robust statistics}

Unless otherwise specified, we measure all global quantities within twice the stellar half-mass radius, $R_\star$, of each galaxy. Quantities related to star formation activity are computed using stars younger than $10 \Myr$. The gas fraction is defined as $\fgas = M_{\rm gas} / (M_{\rm gas} + M_\star)$, where the total gas mass $M_{\rm gas}$ is measured within $2R_\star$ and divided by the total baryonic mass in the same volume. To characterize the size of the star-forming gas phase, we compute $R_{\rm SF}$, the half-mass radius of stars younger than $10 \Myr$, under the assumption that such stars have not significantly migrated from their formation sites over this short timescale.

All statistical analyses presented below are performed using robust estimators suitable for sparse datasets; that is, we adopt the median instead of the mean, and the robust standard deviation (i.e., the median absolute deviation divided by 0.6745) instead of the classical standard deviation \citep[see][for details]{Muller2000, Leys2013, Romeo2023}. We also use robustly standardized quantities, i.e. quantities centered on the median of their distribution and normalized by their robust standard deviation:
\begin{equation}
\label{eqn:std}
	X_{\rm std} = \frac{X-{\rm median}\{X\}}{{\rm median}(\left|X - {\rm median}\{X\}\right|)/ 0.6745},
\end{equation}
which is the robust equivalent of the standardization $(X - \mu)/\sigma$ of a quantity $X$ with mean $\mu$ and standard deviation $\sigma$.

\section{Semantics and star formation regimes}
\label{sec:semantics}

Semantical shortcuts and approximations often made in this field can lead to confusion, misinterpretations, and mistakes. To limit this risk, we list here the definitions adopted throughout the paper, and discuss important subtleties.

By construction, the star formation rate (SFR) combines information on the stellar mass produced and the timescale of production. Therefore, a given SFR can equally represent slow but abundant star formation, or rapid but scarce production. To lift this degeneracy and to characterize the underlying physical processes, one can normalize the SFR by either the stellar mass of the galaxy ($\mstar$), giving the \emph{specific star formation rate}: $\mathrm{sSFR} = \mathrm{SFR}/M_\star$, or by the mass of fuel for star formation ($M_{\rm gas}$), giving the \emph{depletion time}\footnote{The inverse of the depletion time is sometimes referred to as star formation efficiency (SFE). This terminology is problematic because this quantity is not an efficiency (which should be dimensionless) and can be confused with the ratio of stellar mass to gas mass — a true efficiency commonly used at the scale of star-forming clouds. These are distinct quantities with no clear or systematic link between the two. To stop a long-lasting and misleading confusion, we urge the community not to use the terms efficiency or SFE in place of the inverse of the depletion time.}: $\tdep = M_{\rm gas}/{\rm SFR}$. From this, we classify galaxies, as illustrated in \fig{sketch}:

\begin{figure}
\centering
\includegraphics[scale=0.9]{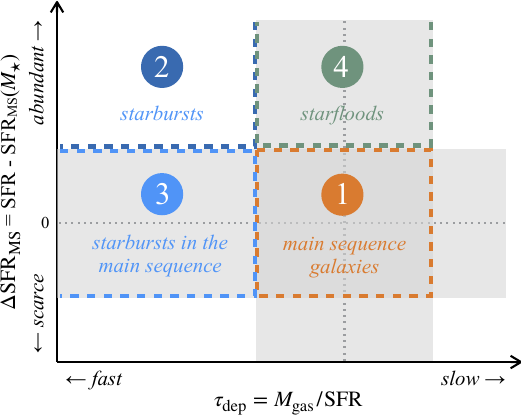}
\caption{Schematic classification of star-forming galaxies based on their timescale of star formation (traced by the depletion time \tdep) and their offset from the main sequence (\dms, i.e., the difference between their SFR and the SFR of the main sequence for their stellar mass). The vertical dotted line represents the median depletion time, and the horizontal one is $\dms=0$. Shaded areas indicate $\pm 1$ times the standard deviation. We highlight four classes and propose an analogy with rivers: the width of the river is \dms, and the speed of the runoff corresponds to the depletion time. A class-1 galaxy (hereafter a main sequence galaxy) has an average SFR for both its stellar and gas masses, analogous to an average river flow. A class-2 galaxy (hereafter a starburst) forms large amounts of stars at an accelerated pace, analogous to rapid and wide waterfalls. As visible in this diagram, the continuous evolution of \dms and \tdep imposes that a galaxy cannot experience a direct transition between classes 1 and 2: it must necessarily pass through classes 3 or 4. Class 3 includes galaxies with rapid star formation but producing average amounts of stars, similar to a tall but narrow waterfall. This work aims to understand the origins of these ``starbursts in the main sequence'' galaxies (SBMSs). Finally, class-4 galaxies produce large amounts of stars but at a normal rate, analogous to floods. We thus introduce the term ``starflood'' to name them.}
\label{fig:sketch}
\end{figure}

\begin{figure}
\centering
\includegraphics{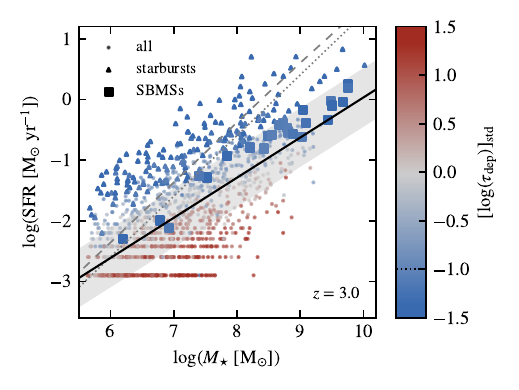}
\caption{Distribution of galaxies in the stellar mass–SFR plane at $z=3$ (for illustration). The color indicates the standardized depletion time (i.e., the difference from the median value of the entire population, divided by the robust standard deviation; see \eqn{std}). The solid line represents the best least absolute deviation fit, and the shaded area indicates $\pm 1$ times the mean absolute deviation from this relation, defining the main sequence of the simulation at this redshift. For comparison, the dotted and dashed lines show the empirical main sequence relations from fits of observed galaxy populations in \citet[their Eq. 9]{Schreiber2015} and \citet[their Eq. 14]{Popesso2023}, respectively (see \citealt{Dubois2021} for a discussion on the main sequence in \nh). The horizontal features at low SFR originate from the finite mass resolution of the simulation.}
\label{fig:ms}
\end{figure}

\begin{figure*}
\centering
\includegraphics{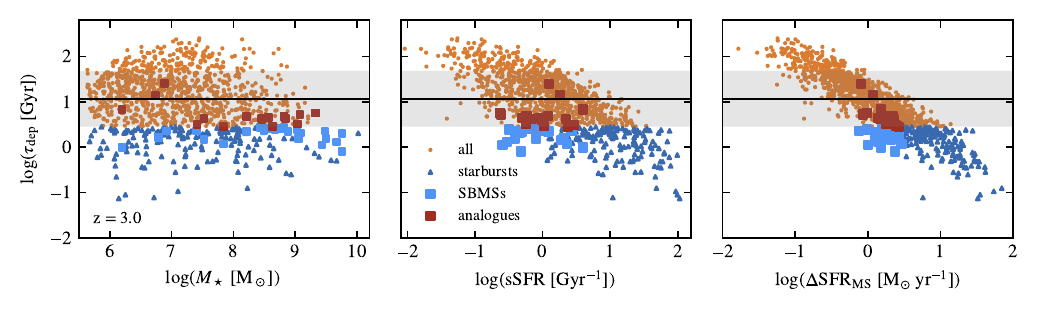}
\caption{Depletion times of the total gas of star-forming galaxies as a function of stellar mass (left) and sSFR (middle) and the offset from the main sequence (\dms, right), shown here at $z=3$. The horizontal line and shaded area show the median \tdep and its dispersion at this redshift. Not surprisingly, the SBMSs are found toward the lower end of the distribution of sSFR of galaxies with short depletion times. This is not a systematic feature since the main sequence is not defined using the sSFR but the best-fit relation in the \mstar–SFR plane.}
\label{fig:ssfr}
\end{figure*}

\begin{itemize}
\item \emph{Main sequence} is the group of galaxies for which the SFR is close to the median of all galaxies of the same stellar mass at the same redshift. The main sequence is an empirical concept that exhibits redshift- and mass-dependences \citep[see][for a connection between the bending mass and the cold-to-hot accretion transition mass]{Daddi2022}. It is defined as a best-fit relation in the \mstar–SFR plane \citep[the form of which remains debated; see, e.g.,][]{Speagle2014,Schreiber2015,Popesso2023}. Here, we consider as members of the main sequence the galaxies with an SFR within $\pm 1 \sigma$ of the best linear fit of the ${\rm log}(\mstar) - \log({\rm SFR})$ relation (which defines $\dms = 0$), derived from all star-forming galaxies at a given redshift. The simulation does not probe the massive end of the stellar mass distribution where the observed main sequence yields an inflection \citep{Noeske2007}, and we discuss this important limitation in \sect{cluster}. Therefore, the functional form we use in the fit to define the main sequence is a simple power-law of the stellar mass, as illustrated in \fig{ms}. The definition of the main sequence changes as galaxies evolve, and thus we re-compute it at every redshift. For convenience, we use hereafter the term ``main sequence'' to refer to galaxies of class 1 only (see \fig{sketch}), spearating the galaxies with a short depletion time (class 3, discussed below).
\item \emph{Starbursts} are galaxies with a short depletion time. In this work, a galaxy is considered a starburst if its standardized depletion time is below minus unity. In other words, a starburst galaxy forms its stars significantly faster than most other galaxies at the same epoch, regardless of the amount of stars formed. Other works use the term starburst to describe sudden and short-lived increases in the SFR, which is fundamentally different from our definition. Following \citet{Renaud2022}, we argue that a definition like ours, based on the timescale of star formation, better reflects changes in the physical regime of star formation, while a definition based on the shape of the SFH could also encompass variations in the availability of star-forming gas without necessarily implying a shift in the mechanisms triggering and/or regulating star formation. Again, for convenience and in agreement with common practice, the term ``starburst'' refers below to class-2 galaxies only (see \fig{sketch}), and excludes the class 3 made of galaxies with low \dms.
\item \emph{Starbursts in the main sequence} (SBMSs) are galaxies fulfilling the criteria of an unusually short depletion time ($\tau_{\rm dep, std} < -1$) and being close to the main sequence ($\dms = 0 \pm 1\sigma$). While these two conditions could seem mutually exclusive at first sight, it is important to note that the depletion time connects the star formation activity (SFR) with its fuel (gas mass), while the main sequence (and the sSFR) relates the SFR to the stellar mass, i.e., the cumulative history of star formation modulated by accretion and stellar mass loss through feedback. Therefore, the depletion time and the offset from the main sequence are fundamentally independent quantities. A galaxy can theoretically be an outlier according to one criterion while being close to the average for the other. In this work, we examine which physical conditions and stages of galaxy evolution lead to such situations. As shown in \fig[s]{ms} and \ref{fig:ssfr} (made at $z=3$ for illustration purposes), SBMS galaxies tend to lie on the upper half of the main sequence and can be found over a wide mass range. However, they are preferentially found toward the high-mass end of the distribution at lower redshifts, as these galaxies acquire their stellar mass faster than most others (see \sect{origins} for more details).
\item \emph{Starfloods} are galaxies above the main sequence (i.e., with $\dms > 1\sigma$) and with a near-median depletion time. As such, their star formation activity is abundant but not particularly fast, hence the analogy with floods. We suspect that such galaxies are rare and/or spend a very short time in this phase during their evolution. We will examine the physics of this regime in a forthcoming paper.
\end{itemize}

In the following, for convenience, the four classes (main sequence, starbursts, SBMSs, and starfloods) are mutually exclusive. We refrain from adopting observationally-derived definitions of the main sequence and the median depletion time, but rather consider the modeled population of galaxies to build consistent definitions. Consequently, we refrain from a direct comparison of our measurements with those from observations, for several reasons. First, our galaxy sample does not probe the high-mass end of the observed distributions, notably where the bulk of SBMSs has been detected in surveys \citep[e.g.,][]{Gomez2022}. This point and a possible extrapolation of our conclusions to this mass range is discussed in \sect{cluster}. Second, not all relevant fundamental physical quantities are easily measurable in observations. For instance, measuring the gas mass requires extensive telescope resources, and is subject to uncertainties araising from dark gas, extinction, and tracers with uncertain conversion factors. Such uncertainties are not systematic, and remain poorly understood \citep[e.g.,][]{Accard2026}. Proxies and alternative, more practical definitions are thus commonly adopted. For instance, galaxies at high redshift are usually classified as starbursts when they exhibit a compact and dust-obscured far-infrared emission \citep[e.g.,][]{Puglisi2021}, consistent with nuclear starbursts often found in local galaxy mergers \citep[but see also][]{Ellison2026}. Such a practical definition is not identical to our physically motivated one and direct comparisons between our work and observational studies cannot possibly be fair. Instead, we compare here the interpretations derived from these works.

In addition, we define the \emph{analogues} as main sequence galaxies with the most similar stellar mass and SFR as the SBMSs: for each SBMS, we identify the nearest\footnote{Since the two axes of \fig{ms} do not share the same dimension, it is not possible to rigorously define an unbiased distance metric in this plane. We nevertheless use a definition based on the coordinate system and units of \fig{ms} to facilitate our interpretations.} point that is a main sequence galaxy in the plane of \fig{ms}. The analogues can then be used as a reference sample in our analysis. With our classification, starbursts share short depletion times with SBMSs, and analogues share similar stellar masses with SBMSs. Therefore, comparing various properties for these classes allows us to identify whether the peculiarities of SBMSs lie in their star formation activity, or the assembly of their stellar mass.

\fig{ssfr} shows the distribution of the galaxies studied at $z=3$ in planes of depletion time, stellar mass, and specific star formation rate. The fundamental difference between \tdep and sSFR noted above explains why these two quantities only very weakly anti-correlate over the entire population, and do not yield any particular relation in the special cases of SBMSs.

\citet{Renaud2022} highlighted the complex, non-trivial and non-monotonous nature of the relation between the position with respect to the main sequence (via \dms or the sSFR) and the depletion time, calling for a deeper analysis as we propose below.

\section{Results}

\subsection{First clues on the nature of the SBMS regime from population statistics}
\label{sec:stats}

We first adopt a statistical approach by monitoring the evolution across cosmic time of the median of key quantities over the entire galaxy populations. This method provides useful indications on the general properties of different families of galaxies. However, these statistics are not representative of individual cases, and thus the connection between such measurements and the underlying physics of galaxy evolution is not direct. Furthermore, the galaxies are re-classified at each snapshot, such that a given class does not track the same subset of objects across time.

The stochasticity of star formation leads to rapid fluctuations in the quantities explored below (see \app{examples} for illustrations). Yet, these fluctuations are not synchronized over the entire galaxy populations, and are thus smoothed out when examining the evolution of median quantities as shown in this section.

\begin{figure}
\centering
\includegraphics{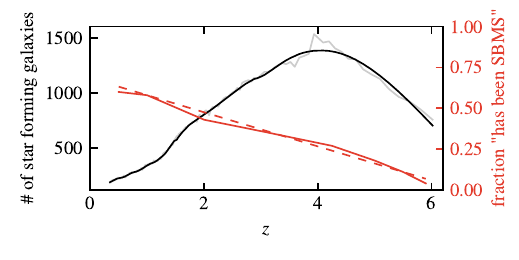}
\caption{Number of star-forming galaxies considered in our sample as a function of redshift (grey line and its smoothed version in black, left axis), and fraction of these galaxies that have experienced at least one passage in the SBMS regime in their lifetime (red solid line, and linear fit in dashed red, right axis).}
\label{fig:number}
\end{figure}

\fig{number} shows that the fraction of star-forming galaxies that have experienced at least one SBMS episode in their lifetime is a linear function of redshift: $-0.10z + 0.68$: after cosmic noon, the majority of star-forming galaxies have been through the SBMS regime at least once. Although this best-fit relation predicts the onset of SBMS galaxies at $z \approx 6.5$, the earliest occurrence in our sample is found at $z = 5.9$. This is, by construction, subject to low-number statistical effects.

\begin{figure}
\centering
\includegraphics{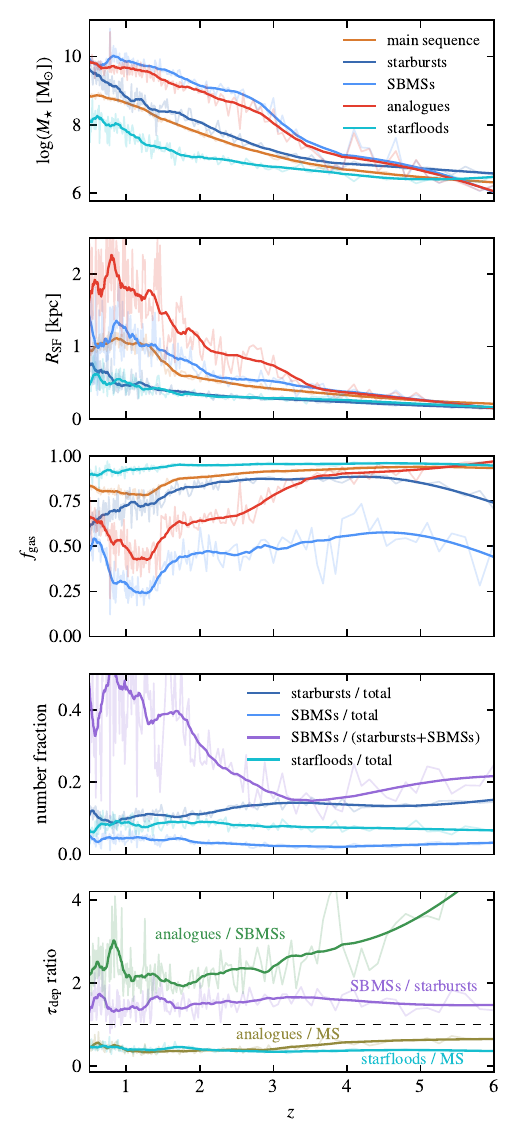}
\caption{\emph{Top three panels:} evolution of the median of the logarithmic stellar mass, the half-mass radius of the star-forming gas, and the gas fraction for galaxies in the main sequence, in the starburst regime, in the SBMS regime and their analogues (i.e., the non-starburst main sequence galaxy with the most similar SFR and \mstar), and in the starflood regime (see \sect{semantics} for details). \emph{Penultimate panel:} evolution of the number fractions of each regime over the entire population of star-forming galaxies. Also shown is the fraction of SBMSs among galaxies with short depletion times (i.e., starbursts and SBMSs). \emph{Bottom panel:} median depletion time of the SBMSs normalized by that of the starburst galaxies, and the same ratio between the analogues and the main sequence galaxies, and between the analogues and the SBMSs. The horizontal line indicates unity. All curves have been smoothed using a Savitzky–Golay algorithm to improve readability, with the original measurements shown by the semi-transparent lines.}
\label{fig:redshift}
\end{figure}

The top panel of \fig{redshift} reveals that the SBMSs and their analogues are almost always (on average) more massive than the main sequence and starburst galaxies. However, the size of the star-forming gas component (second panel) is significantly smaller in SBMSs than in galaxies with similar SFR and stellar mass. This could be explained by a precocious assembly of the stellar mass of the SBMS galaxies, followed by a slowing down of their star formation activity, possibly due to a dilution of the gaseous phase (see \sect{episodes}). It may thus be possible --but not inevitable-- that SBMS were previously starburst galaxies which have returned to the main sequence, as examined in \sect{origins}. Similarities in median stellar mass between the SBMSs and the analogues at all epochs directly reflect the selection criteria adopted. However, it should not be interpreted as analogues having the same stellar mass assembly as the SBMSs. This difference is further explored in \sect{origins} where we find that SBMSs have experienced more frequent and efficient star-forming merger events than any other category. At all redshifts, the SBMSs have a significantly lower gas fraction (third panel) than any other class --as expected, on average, from the tightness of the main sequence-- and supporting the hypothesis of an overly massive stellar component. 

The penultimate panel of \fig{redshift} shows that the evolution of the number fraction of star-forming galaxies in the starburst and SBMS classes is mild (between 9 and 15\% for starbursts, and between 3 and 6\% for the SBMSs). This panel also shows that the fraction of SBMS galaxies among the population of galaxies with short depletion times (starbursts and SBMSs) remains approximately constant at $13-18\%$ until $z \approx 2.5$, and increases to $\gtrsim 30\%$ for $z \lesssim 2$. Therefore, the physical conditions necessary for a galaxy to become SBMS change only mildly across cosmic time, at least not monotonically. 

Finally, the bottom panel of \fig{redshift} shows that the median depletion time of the SBMSs is $\approx 1.4$ times longer than that of starbursts, with no significant variation with redshift. While significant, this difference does not necessarily imply a distinct star formation regime in SBMSs compared to starbursts. In \sect{origins}, we show that SBMSs experience more frequent and productive merger events, likely caused by their earlier and faster formation. This translates into shorter depletion time than the average starburst, as visible here, without suggesting a discontinuity in the star formation regime. It also indicates that, on average, SBMSs do not constitute a transition regime lying between starsbursts and main sequence galaxies. Thus, SBMSs are not merely the high-stellar-mass end of the distribution of starburst galaxies, which would naturally explain their lower sSFR, but may instead correspond to a different stage of their evolution. Conversely, the depletion time of the analogues is $\approx 0.5$ times shorter than that of the main sequence galaxies, and the SBMSs exhibit depletion times at least twice as short as their analogues. 

\begin{figure}
\centering
\includegraphics{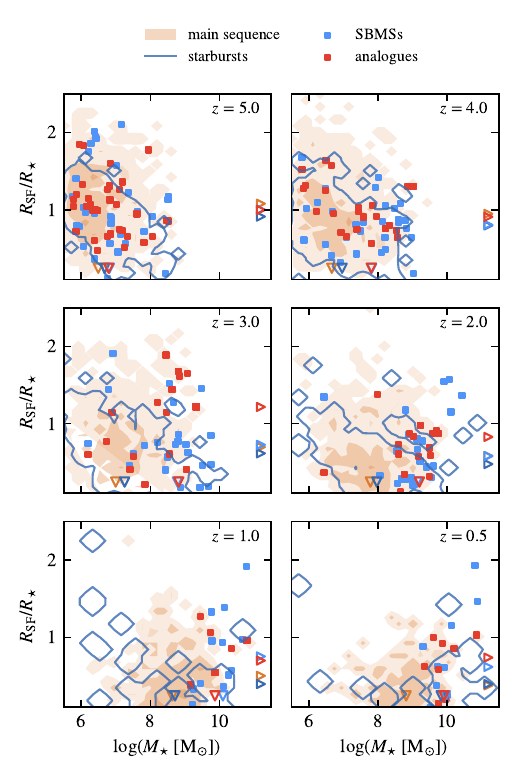}
\caption{Ratio of the size of the star-forming ($R_{\rm SF}$) to the stellar component of galaxies, as a function of stellar mass, for all star-forming galaxies at six different redshifts. $R_{\rm SF}$ is computed as the half-mass radius of the stars younger than $10\Myr$, and is used as a proxy for the size of the star-forming gas phase. Filled contours indicate the distribution of the main sequence galaxies, and that of the starbursts is shown by the solid line contour. The SBMSs and their analogues are shown as squares. Triangles on the right and bottom axes indicate the median values of both quantities for each population. For $z \gtrsim 2.0$, SBMS galaxies are found across the mass and size ranges of the star-forming galaxies and are only confined to the high-mass end at later epochs. Apart from this, the distributions of SBMS galaxies in this plane are not very different from those of the other classes, which contradicts scenarios in which a compaction event marks the onset of the SBMS regime.}
\label{fig:masssize_ratio}
\end{figure}

\begin{figure}
\centering
\includegraphics{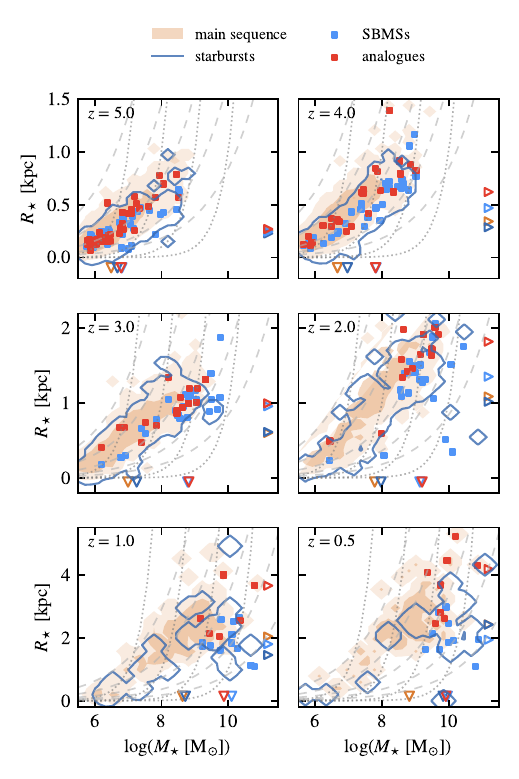}
\caption{Same as \fig{masssize_ratio} but showing the stellar mass–size relation. Dotted (respectively dashed) guiding lines indicate constant compactness ($M_\star \propto R_\star$) and constant density ($M_\star \propto R_\star^3$), spaced by factors of 10 and 2, respectively. The range of the vertical axis changes between the rows for readability, but the guiding lines remain identical in all panels. On average, the SBMSs are more compact than galaxies from the other classes.}
\label{fig:masssize}
\end{figure}

Several scenarios proposed to explain the SBMS regime invoke a compaction event in the form of a rapid shrinking of the size of the star-forming material. To test this hypothesis, \fig{masssize_ratio} displays the size ratio of the star-forming to stellar components. The absence of a clear separation of the SBMSs from the other galaxies indicates that their star-forming material does not experience a significant change in size \emph{with respect to the stellar component}. Still small but more significant differences appear in \fig{masssize}, where most SBMSs have a more compact stellar component than that of main sequence galaxies and also in starbursts (but only a few percent more compact than their non-starbursting analogues). However, this trend is not visible when comparing the median size of the classes, i.e., when marginalizing over the stellar mass. The persistence of this distinction at all redshifts suggests that the effect is physical and not due to small-number statistics. Thus, the compact nature of these galaxies concerns both the stellar and dense gas components, implying a dynamical origin (e.g., tidal stripping of the outer galaxy, loss of angular momentum from tidal torques) but rejecting the possibility of a hydrodynamical-only process which would only affect the gas component (e.g., loss of momentum through shocks or ram pressure). Other configurations might be possible when considering denser environments like that of galaxy clusters, which are not probed in the volume of the \nh simulation.

This contradicts one of the interpretations of \citet{Magnelli2023} and \citet{Lyu2025}, who, using the CEERS survey (JWST/MIRI), identify galaxies with a compact star-forming component embedded within an extended stellar disk (their ``EC'' class) and tentatively associate them with SBMS. In their scenario, the EC/SBMS event is triggered by the compaction of the dense gas, increasing the central star formation density, and driving the in-situ build-up of a bulge-like morphology (see also \citealt{Lebail2024, Tan2024} reporting centrally luminous infrared galaxies with extended stellar disks). In our simulation, such an evolution also triggers an elevation of the star formation rate, but not a drop in depletion time, and thus cannot be associated with the SBMS regime (see \fig{growth} for details). The discrepancy originates from the different definitions used to identify starbursts. While we base our criterion on physical quantities, such measurements are notoriously more challenging in observations. The gas mass is often unavailable, or at best subject to important uncertainties and the necessity to assume standard scaling relations to derive it \citep[e.g.,][]{Accard2025}. As a result, the observed galaxies are rarely classified according to their depletion time but by the compact nature of their star-forming component (as in \citealt{Magnelli2023, Lyu2025}), which blurs or skews their identification as starbursts or SBMSs.

Our results are also in tension with the scenario of \citet{Puglisi2021}, who propose that SBMSs result from merger-driven nuclear gas inflows leading to starburst activity, followed by a post-merger evolution through the main sequence and toward quenching \citep[e.g.,][]{Barro2014, Puglisi2019}, but maintaining an extended stellar component \citep[see also][]{Elbaz2018, Franco2020, Tadaki2020}. During a merger, tidal torques affect all components \citep{Mihos1996} and thus, purely gaseous inflows could only be caused by another mechanism as discussed in \citealt{Tan2024} to explain the in-situ build-up of bulge-like structures in submillimeter galaxies. 

The picture emerging from our analysis so far is that of an early and rapid assembly of the stellar component favoring a low sSFR to explain the properties of the SBMSs. We stress that this rapid assembly does not necessarily immediately precede the SBMS episode (see \sect{episodes} for an analysis of the short-term evolution of individual galaxies before and after the SBMS phase). Contrary to interpretations from the literature, we do not detect in SBMSs any significant recent compaction of the gas reservoir in an extended stellar component (i.e., SBMSs, whose starburst nature is defined by depletion time rather than a compact infrared component, are not necessarily associated with the EC galaxies of \citealt{Lyu2025}) but rather both stellar and gaseous components being more compact than average (the CC class of \citealt{Lyu2025}). We discuss in \sect{episodes} whether the compactness is a fundamental trigger of the SBMS regime. Furthermore, the peculiarities of the SBMSs do not seem to vary with redshift.

We cannot yet conclude on the role of punctual events such as mergers, as their impacts are not synchronized over the full population of galaxies and are thus smoothed out in statistical analyses. Two examples of the full history of SBMS galaxies at $z=3$ are shown in \app{examples} to illustrate the complex and non-monotonic evolution of galaxies above, within, and below the main sequence. All cases we examined yield similar highly-time-varying, stochastic, evolutions (see also \citealt{Gui2025}). This contradicts a number of proposed scenarios in which the SBMS phase is a mere transition between a starburst phase and definitive quenching \citep[e.g.][see also \sect{episodes}]{Tacchella2016, Puglisi2019}. We stress again that the fluctuations noted in these examples are smoothed out when computing statistical quantities over a large number of galaxies (e.g., the main sequence), demonstrating that statistics are not representative of individual histories, and thus cannot probe full complexity of the underlying physical processes.

\subsection{Early history of SBMS galaxies: the role of mergers}
\label{sec:origins}

The results of the previous section suggest that at least part of the peculiarities of SBMS galaxies originates in a precocious assembly of their stellar component, from external (accretion of companion galaxies) and/or in situ origins (star formation). To explore this further, we compare the histories of SBMSs with those of the other classes of galaxies. To ease interpretation, we focus at $z\approx 3$ as a compromise between an epoch late enough to cover a diversity of galactic morphologies and histories \citep[see][]{Dubois2021, Kraljic2024} and still early enough to keep their histories simple. We generalize our conclusions to other redshifts in \sect{redshift}. The galaxies are classified at $z=3$ according to the definitions in \sect{semantics}, and then tracked back in time by identifying their most massive progenitor on each previous snapshot, i.e., every $\approx 10\mh 15 \myr$.

\begin{figure}
\centering
\includegraphics{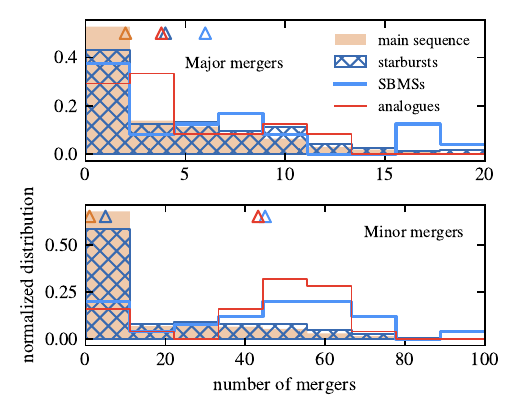}
\caption{Normalized distributions of the number of major (top, mass ratio $>$ 1:10) and minor mergers (bottom, mass ratio between 1:10 and 1:50) experienced by galaxies identified and classified at $z=3$. The triangles indicate the median values for each population. Changing the mass-ratio thresholds for the definitions of major and minor mergers alters these distributions but not our conclusions. SBMS galaxies experience more major and minor merger events than the other classes, which contributes to the early build-up of their stellar mass by accretion and merger-triggered starbursts (see also \fig[s]{timeInStarburst} and \ref{fig:medianDepletionTime}).}
\label{fig:nmergers}
\end{figure}

\fig{nmergers} reveals that the median number of major mergers in the history of SBMS galaxies is 1.5 times higher than for the starbursts and the analogues. Despite SBMSs being preferentially found among the most massive galaxies (which implies most of their encounters happen with lower-mass galaxies), they experience significantly higher numbers of both minor and major mergers than the starburst galaxies, and more major mergers than their analogues. Therefore, the increased stellar mass of SBMSs is caused, in part, by repeated merger events.

\begin{figure}
\centering
\includegraphics{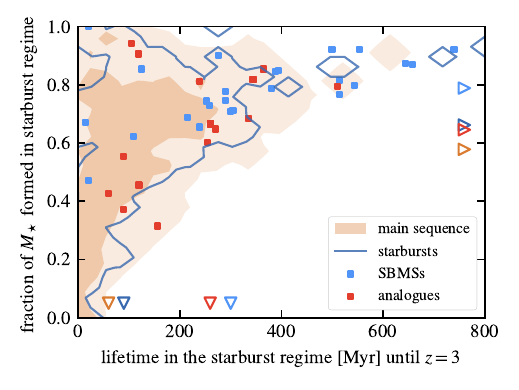}
\caption{Importance of the starburst regime in the history of the galaxies found at $z=3$, shown by the lifetime they spent as starburst and the fraction of their stellar mass they formed in this regime (irrespective of the number of starburst episodes experienced). The triangles near the bottom and right axes indicate the median values for each population. SBMSs are clearly distinguished by longer and more productive starburst episodes.}
\label{fig:timeInStarburst}
\end{figure}

\begin{figure}
\centering
\includegraphics{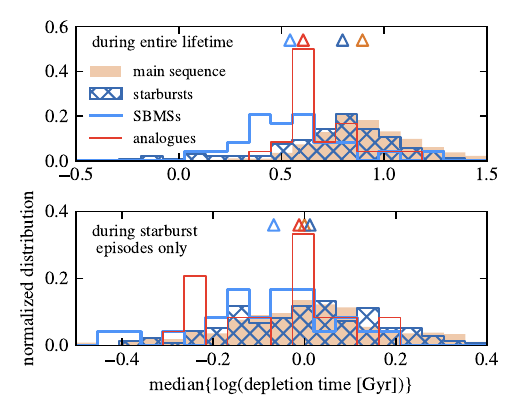}
\caption{\emph{Top:} Normalized distributions of the median logarithmic depletion time of galaxies between their formation and their classification at $z=3$. \emph{Bottom:} same but considering only the depletion times during the starburst episodes. Triangles indicate the median of the distributions.}
\label{fig:medianDepletionTime}
\end{figure}

In addition to accretion of neighbor galaxies, in situ star formation is the only other pathway to grow the stellar component. Rapid star formation is often associated with wet major mergers which can trigger nuclear inflows, shocks, and tidal and turbulent compression. Experiencing many major mergers is thus compatible with a precocious growth of the stellar mass from both in situ and external origins. However, the response of galaxies to the stimulus of a major merger in terms of star formation is not constant across cosmic time as it depends on the morphology \citep{Segovia2022} and gas fraction of the progenitor galaxies \citep{Perret2014, Fensch2017}, and possibly on other factors too. 

To establish the role of mergers in the build-up of stellar mass, one must account for all past mergers and not only those in the most recent history, as commonly done when linking star formation to starburst activity. Therefore, in what follows, we consider all merger events experienced by the galaxies up to the epoch of interest ($z=3$ in our case). Although examining each merger event individually is beyond the scope of this paper, \fig{timeInStarburst} shows that the SBMSs and their analogues have spent respectively 3.5 and 4 times more time in the starburst regime than the main sequence and the other starburst galaxies at $z=3$.\footnote{With the cosmological parameters assumed in \nh, the age of the Universe is $\approx 2200 \myr$ at $z=3$.} The difference between the SBMSs and their analogues is more pronounced when considering the fraction of stellar mass formed during these starburst episodes: SBMSs stand out by a production of stars 20\% faster than the other galaxies including their analogues, that is, a short depletion time during the starburst episodes they experienced, as confirmed in \fig{medianDepletionTime}. When considering their entire lifetime (top panel of \fig{medianDepletionTime}), SBMSs and their analogues have similar median depletion times, about 1.6 times shorter than the starbursts and main sequence galaxies. However, the distribution of SBMSs extends significantly further toward short depletion times, and episodes of long depletion times are rarer for SBMS galaxies than for the others. This could potentially be explained solely by the differences in number of major mergers (\fig{nmergers}), but when selecting only the starburst episodes along the lifetime (bottom panel), the SBMSs appear to host star formation approximately 15\% faster than that of all the other galaxies. Furthermore, major mergers of starburst and SBMS galaxies are 1.4 times more frequent than for galaxies with longer depletion times (\fig{MMfrequency}) and the ratio of major to minor mergers is 1.3 times higher for SBMSs than for their analogues. Owing that interactions with massive galaxies are generally more efficient at triggering starbursts than those involving a smaller companion \citep{Cox2008, Ellison2008}, mergers that are more numerous, more frequent, and involve more massive companions explain the rapid star formation in SBMSs.

For the events considered here, we find no particular relation between short depletion times and an early settling of the disk (which would be more prone to responding to mergers by triggering starburst events; \citealt{Segovia2022}), nor with a lower gas fraction than in other galaxies \citep{Perret2014, Fensch2017}. The underlying physical cause of the merger-triggered starbursts could then possibly be stronger compression generating an excess of dense gas which, in turn, increases the SFR and decreases the depletion time \citep{Renaud2014b}. Repetitions of such events enable these galaxies to quickly assemble their stellar component, in addition to the frequent accretion of their companions, which naturally tends to lower their sSFR and \dms, and make them SBMSs. \citet[their Figure 3]{Renaud2022} shows that the compression varies strongly with the large-scale environment, suggesting that SBMSs could lie in specific locations of the cosmic web favoring frequent and starburst-efficient mergers. We will test this hypothesis in a forthcoming paper.

In summary, on average, SBMS galaxies distinguish themselves by precocious and intense starburst activity likely connected to frequent major mergers, during which they assemble a significant fraction of their stellar component by accretion and merger-triggered rapid star formation. Galaxies from the other classes experience fewer and less frequent mergers, spend less time in starburst mode, and yield longer depletion times during these episodes. The dispersions of the distributions shown in \fig[s]{timeInStarburst}, \ref{fig:medianDepletionTime}, and \ref{fig:MMfrequency} suggest that these generalities cover a diversity of finer scenarios, which we examine further in the next section.

\subsection{Evolution before, during, and after the SBMS episodes}
\label{sec:episodes}

\fig{sketch} shows that the SBMS class can be a transition stage between starburst and main sequence, as often supposed in the literature. However, other paths are possible in the \tdep--\dms plane. In this section, we classify SBMS galaxies according to their class immediately before and immediately after their SBMS episode. Four sub-categories of SBMS\footnote{The transition of SBMS galaxies to/from galaxies with short depletion times and low \dms (i.e., ``starbursts \emph{below} the main sequence'') is not detected in the simulation.} appear according to the chronological sequence of the classes:
\begin{itemize}
\item 1-3-2: from main sequence to starburst (``MS to SB''),
\item 2-3-2: from starburst to starburst (``SB to SB''),
\item 2-3-1: from starburst to main sequence (``SB to MS''),
\item 1-3-1: from main sequence to main sequence (``MS to MS'').
\end{itemize}
To improve the statistics\footnote{Cases for which the phase immediately preceding and/or succeeding the SBMS episode lasts less than two consecutive snapshots are ignored. This removes the SBMSs and analogues of the lowest mass ($\lesssim 10^7 \msun$) from the sample considered in this section, as the extreme stochasticity of their star formation properties challenges a robust identification of their evolutionary path through the SBMS phase.}, we stack the SBMS episodes detected on snapshots within $\pm 250 \Myr$ of $z=3$. In this time window, the four sub-classes comprise approximately $9\%$, $39\%$, $22\%$, and $30\%$ of the SBMS galaxies, respectively. Therefore, about two-thirds of the galaxies return to their previous regime after the SBMS phase. (These numbers vary with redshift; see \sect{redshift}.) Among the remaining third, the majority reduces its star formation activity (``SB to MS''). Despite being the most discussed option in the litterature \citep[as a possible route toward quenching; see e.g.,][]{Puglisi2021, Gomez2022}, this pathway concerns less than one quarter of the entire SBMS population in \nh.

\begin{figure}
\centering
\includegraphics{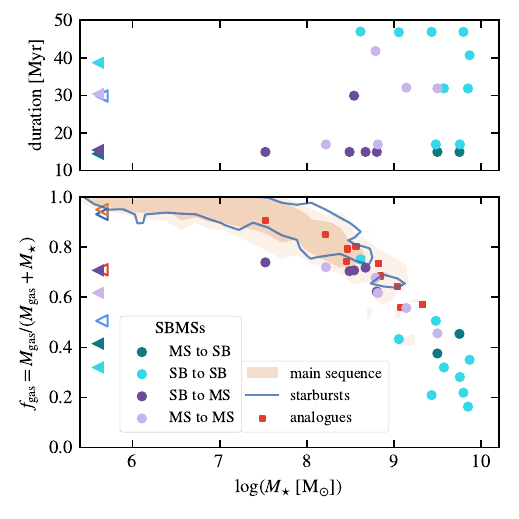}
\caption{Duration of the current SBMS episode (\emph{top}) and gas fraction (\emph{bottom}) for the SBMS galaxies detected at $z=3$, shown as a function of stellar mass. The SBMS galaxies are classified according to their trajectory in the \tdep \mh \dms plane before and after the SBMS phase (see text). The triangles to the left of the panels indicate the median value of each group (with all SBMSs in light blue). In the top panel, a small vertical offset has been added to the ``SB to SB'' and ``MS to MS'' points to avoid graphical overlap with other points; this offset is not used when computing the medians.}
\label{fig:duration}
\end{figure}

\begin{figure*} 
\centering
\includegraphics{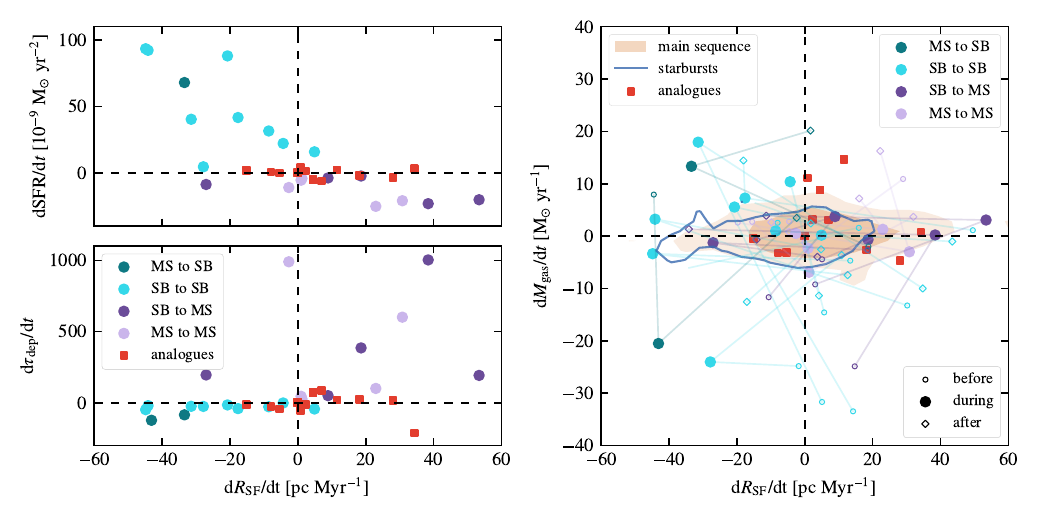}
\caption{Variations of the star formation rate (\emph{top-left}), of the depletion time (\emph{bottom-left}), and of the gas mass within twice the half-mass stellar radius (\emph{right}), compared to the change in half-mass radius of the star-forming gas (traced by stars younger than $10 \Myr$) at $z=3$. All variations are measured over the exact duration of each galaxy's SBMS episode, or over $30 \Myr$ for the non-SBMS galaxies shown for reference (i.e., similar to the median duration of an SBMS episode; recall \fig{duration}). SBMSs are color-coded according to the sub-class they belong to before and after the SBMS phase (see text). The main sequence and starburst galaxies are omitted in the left panels for clarity. In the right panel, for each SBMS galaxy, lines connect the measurement during the SBMS episode to measurements $30 \Myr$ immediately before (empty circles) and after (empty diamonds) the episode, indicating a three-point evolutionary track in this plane.}
\label{fig:growth}
\end{figure*}

\fig{duration} shows that, at $z=3$, the longest SBMS episodes occur in the most massive galaxies, and most of these massive systems then return to their previous class (``SB to SB'' and ``MS to MS''). Shorter SBMS episodes occur across all masses and evolutionary types. Rapid in-and-out transitions of the SBMS phase are likely caused by fluctuations of the SFR on comparable timescales (see \fig{track}). Lower-mass galaxies, being more responsive to rapidly changing internal and environmental regulators of star formation (feedback, compaction, mergers; \citealt{Dekel2014, Muratov2015, Christensen2016}), naturally experience the shortest SBMS episodes. Galaxies that transition from one regime to another (``MS to SB'' or ``SB to MS'') spend the least time as SBMS. The limited output cadence of the simulation (of order $15 \Myr$ at this epoch) likely influences these measurements: the shortest SBMS episodes may be unresolved.

The bottom panel of \fig{duration} confirms the peculiarity of SBMSs and their sub-classes in terms of gas fraction. Overall, the gas fraction of our star-forming galaxies decreases monotonically with increasing stellar mass. Regardless of mass, SBMSs lie at the lower end of the gas-fraction distribution. These low \fgas mainly result from a high stellar mass (which we also invoke to explain their low sSFR and low \dms, recall \sect{stats}) rather than small amounts of gas. Therefore, even with low \fgas, SBMSs are potentially capable of re-igniting a starburst activity. Indeed, the most massive SBMSs --those with the lowest gas fractions-- are galaxies on the verge of becoming starbursts within, at most, a few $10 \Myr$ (as noted above). At slightly lower masses --and thus significantly higher gas fractions-- lie SBMSs that will next become main sequence galaxies. Therefore, the short-term future evolution of SBMSs is related to the gas fraction. Conversely, their recent past does not correlate with either gas fraction or stellar mass, since these quantities result from the assembly of the gas and stellar components over long timescales.

Several scenarios propose to connect SBMSs to a compaction episode that would trigger starburst activity in turn leading to the exhaustion of the star-forming gas \citep[e.g.,][]{Gomez2022}. However, the link between kpc-scale accumulation of gas and the pc-scale change of gas density that sets the star formation activity is not yet fully understood. In particular, while a compaction event is commonly expected to increase the SFR by raising the amount of dense gas, it is not clear whether this necessarily implies a decrease in depletion time \citep{Lapiner2023}. In any case, our analysis does not confirm a dominant role of compaction in the SBMS phase (recall also \fig{masssize_ratio}), but the statistical nature of the results shown in \sect{stats} leaves open the possibility that this scenario only concerns a specific evolutionary pattern limited to a sub-category of SBMSs, as we test below.


The left panels of \fig{growth} compare variations in the size of the star-forming ISM (traced by the half-mass radius of stars younger than $10 \Myr$) to changes in SFR and depletion time over the entire SBMS episode, and over $30 \Myr$ for the analogue galaxies shown for reference. These analogue galaxies maintain their SFRs and depletion times despite non-negligible compaction or dilation of their star-forming ISM (up to kpc scales during the period considered). This strikingly contrasts with the SBMS galaxies, which show unambiguous responses to size variations. Moreover, a clear dichotomy appears in this figure between SBMSs that become starbursts and those that become main sequence galaxies.

SBMSs that will become starbursts (``MS to SB'' and ``SB to SB'') experience a compaction of their star-forming ISM ($\dd R_{\rm SF} / \dd t < 0$). This translates into an increase of the SFR (approximately as $\dd {\rm SFR} / \dd t \propto -1.5 \msun \kpc^{-1} \yr^{-1}\, \times\,  \dd R_{\rm SF} / \dd t $), but only into a very weak acceleration of the star formation process. Indeed, the variations in \tdep for these galaxies are of the same order as those in the analogue reference sample, and thus not related to the SBMS physics. Conversely, SBMSs on their way to become main sequence galaxies (``SB to MS'' and ``MS to MS'') experience a dilation of their star-forming ISM during the SBMS episode ($\dd R_{\rm SF} / \dd t > 0$), which induces a mild diminution of the SFR (yet significantly more pronounced than the variations in the analogue sample). Given the moderate change in gas mass for these galaxies (visible in the right panel of \fig{growth}), this necessarily leads to a lengthening of the depletion time, as confirmed in the bottom-left panel (approximately $\dd \tdep / \dd t \propto 4.2 \Myr \msun \pc^{-1}\, \times\, \dd R_{\rm SF} / \dd t $, but with an important scatter). Such dilation events are thus directly connected to the change of star formation regime from starburst (abundant and fast) to main sequence (scarcer and slower). If caused by feedback, the absence of significant mass-loss indicates a redistribution of the gas within the galaxy instead of outflows, or a balance by accretion at the same rate.

The right panel of \fig{growth} also shows the evolution of galaxies shortly before, during, and shortly after their SBMS phase at $z=3$. The variety of tracks contrasts with the simplicity of statistical trends described above: while SBMS galaxies appear as systems that precociously assemble their stellar mass (\sect{origins}), their evolution over short-timescales ($\sim 30 \mh 100\ \Myr$) is far from universal. Nevertheless, all tracks reach an extremum in either mass- or size-growth rate (or both) during the SBMS phase, confirming the peculiar and transitional character of this phase along galaxy evolution. It appears that the nature of this extremum and its relation to the preceding and following phases distinguish the SBMS sub-classes.

Starbursts that experience an SBMS phase before returning to the starburst class (``SB to SB'') typically show an initial reduction of gas mass due to accelerated consumption by star formation, accompanied by a mild size growth of the star-forming phase --possibly caused by efficient gas redistribution by stellar feedback. After a compaction during the SBMS phase (often associated with an accumulation of gas), they resume their off-main-sequence starburst activity with a size expansion but with no unique trend in the evolution of their gas mass. This diversity likely reflects the variety of physical processes that trigger starbursts, each with its own timescale that can or cannot overcome the replenishment of the gas reservoir \citep{Renaud2019b}. For the same reason, no clear trend is found for main sequence galaxies that become starbursts (``MS to SB''). In virtually all cases, galaxies that become starbursts after the SBMS phase experience the strongest compaction of their star-forming ISM during the SBMS phase, along the scenario of formation of blue nuggets \citep{Dekel2014}. However, because the SBMS phase is short-lived (\fig{duration}), these galaxies do not have enough time to exhaust their gas reservoir and do not become red nuggets. Therefore, compaction leading to a SBMS phase is not a direct pathway toward a quiescent or quenched galaxy, but rather toward the (re-)ignition of a starburst activity. 

Main sequence galaxies that temporarily become SBMSs (``MS to MS'') do not experience statistically significant changes in gas mass or size: their distributions in the right panel of \fig{growth} are concentrated around the no-growth position, similar to the non-SBMS galaxies. The brief passage through the SBMS phase for these systems could be caused by rapid fluctuations in the conditions for star formation. More generally, the gas mass of SBMS galaxies that move toward the main sequence (``MS to MS'' and ``SB to MS'') does not change significantly during the SBMS episode, but tends to increase slightly afterward. This could possibly result from an imbalance between steady gas inflows from large scales and a slowed consumption by star formation. Thus, whether an SBMS progenitor was a starburst or a main sequence member, an SBMS episode does not mark a step toward quenching.

The evolutionary path from starburst to quiescent galaxy via the SBMS regime, as proposed by \citet{Puglisi2021} in the context of quenching, concerns at most a quarter of our simulated sample, and this fraction is even lower at other redshifts, see \sect{redshift}. Our measurements rather support scenarios where the SBMS is a short-lived step along a diversity of evolutions, better compatible with the first steps in the proposition of \citet{Gomez2022}, namely repeated passages in and out of the main sequence, before quenching \citep[see also][]{Lyu2025}. Adopting different definitions of the galaxy classes and characterizing the star formation activity over longer timescales (as done when deriving the SFR from UV light for instance) would likely change our quantitative results. Yet, the fact that our approach does capture the SBMS regime across the full diversity of possible paths demonstrates the robustness of our qualitative conclusions. We discuss the sensitivity of our conclusions to the choice of numerical method and initial conditions, notably the galactic mass, in \sect{discussion}.

\subsection{Generalization to other redshifts}
\label{sec:redshift}

The conclusions above have been obtained from the analysis of the galaxy population at $z=3$. Generalizing to higher redshifts suffers from low-number statistics, while at later epochs, galaxies experience increasingly complex formation histories, possibly complicating the interpretations proposed above. Nevertheless, both the cosmological environment (e.g., the merger rate) and the intrinsic evolution of galaxies require assessing the validity range of our conclusions. We describe here the main differences between our conclusions at $z=3$ and results from the same protocol conducted at other redshifts\footnote{As before, the main sequence and median quantities are re-computed and the galaxies are re-classified at each redshift.}.

The most prominent evolution with redshift is the progressive depletion of SBMS galaxies toward the low-mass end of the galaxy mass function at low redshift (visible in, e.g., \fig{masssize_ratio}). This naturally derives from the early assembly of the stellar component via accretion and enhanced in situ star formation, which is central to our scenario.

At high redshift ($z\gtrsim 4$), despite lower number statistics, virtually all the quantities examined yield smaller scatters. This is caused by galaxies having had less time to significantly diverge from a relatively uniform evolutionary pathway, i.e., the diversity of histories is not yet fully developed. Considering a larger cosmological volume (e.g., including galaxy protoclusters) would likely mitigate this conclusion.

Towards low redshifts, the number of mergers experienced by SBMSs increases less than that of any other class: after $z\approx 1$, the other classes yield a higher number of major mergers, and a similar number of minor mergers than SBMSs, on average (see \fig{nmergers_z1}). The frequency of major mergers steadily decreases with time for all classes \citep{Martin2021}, and ceases to be enhanced for SBMSs at $z \lesssim 1$ (\fig{MMfrequency_z1}).

Although the median duration of the SBMS episodes does not change across cosmic time, short episodes ($\lesssim 20 \Myr$) become rarer at low redshifts. This is likely caused by longer dynamical timescales in increasingly more massive galaxies. Notably, the ``MS to SB'' transition is twice as long at cosmic noon ($z\approx 2$) as at any other redshift ($\approx 40\Myr$), possibly linked to a longer interaction timescales if this transition is merger-triggered. The fraction of SBMS galaxies returning to their previous class (``SB to SB'' and ``MS to MS'') increases significantly with decreasing redshift: while these sub-categories are absent at $z\gtrsim 3$, they represent $\approx 70\%$ of the SBMSs at $z \lesssim 2$. The transition from starburst to main sequence (``SB to MS'') is virtually nonexistant at $z\ge 4$ and remains rare at all redshifts ($2\%$ at $z=2$, $10\%$ at $z=1$), with its maximum reached at $z=3$ ($22\%$). This further challenges the proposition commonly found in the literature of SBMS being a transition phase between starburst and quiescent galaxies: while possible, this scenario concerns less than a quarter of the SBMS galaxies at any redshift.

The dependences of the variations of SFR, \tdep, and the size of the star-forming component remain the same at all redshifts, with the same differentiation between the sub-categories of SBMS galaxies. We note, however, that variations in \tdep during the SBMS phase are less important at both higher and lower redshifts than at $z=3$. 

In summary, the scenario proposed in the previous sections from our analysis at $z=3$ remains valid throughout cosmic time. Quantitative variations arise from the overall evolution of the gas contents and merger rate, but follow the general trends of galaxy evolution, irrespective of the SBMS regime. This demonstrates that the SBMS behavior is a recurrent mode of star formation, embedded within the broader evolutionary trends of star-forming galaxies.

\section{Discusion}
\label{sec:discussion}

\subsection{Extrapolation to more massive galaxies}
\label{sec:cluster}

In the GOODS-ALMA sample, SBMSs are found between $\approx 10^{10.2}$ and $10^{11.5} \msun$ at $z\approx 1\mh 2$ \citep{Elbaz2018, Ciesla2023}, that is, approximately an order of magnitude more massive than the average of our modelled cases (\fig{redshift}). The \nh simulation has been designed to explicitly avoid dense cosmological environments favoring the assembly of the most massive galaxies, such as galaxy clusters. As a result, the maximum stellar mass in the simulation volume is $\mstar \approx 10^{11} \msun$ at $z=0$. Since different mechanisms are involved in galaxy evolution across the mass spectrum, the mass discrepancy between the observed and modelled samples could indicate that our scenario for SBMSs might not fully apply to more massive galaxies.

Although the merger rate only weakly scales with stellar mass \citep{Lotz2011, OLeary2021}, other mechanisms show a much stronger mass dependence. For example, gas accretion transitions from cold to shock-heated at virial masses of $\approx 10^{12} \msun$ \citep{Birnboim2003}. Combined with stronger AGN feedback and shorter depletion times \citep{Fabian2012, Tacconi2020}, this can result in a slower replenishment of the gas reservoir after an intense starburst episode \citep{Dekel2006, Peng2010}. In this case, the SBMS phase would then represent a transition toward quenching, as postulated by \citet{Tacchella2016} and \citet{Gomez2022}. 

However, our analysis reveals very few and only weak dependencies of our scenario on stellar mass. Therefore, either our conclusions apply across the mass spectrum, or there exists a critical mass (such as the $\approx 10^{12} \msun$ threshold for the virial mass from \citealt{Birnboim2003} and proposed by \citealt{Daddi2022} as a transition to explain the bending of the main sequence as a fonction of stellar mass; see also the discussion in \citealt{Lapiner2023} on its role on the compaction mechanism) above which our scenario must be adjusted to account for different relative roles of the physical processes governing galaxy evolution. For instance, \citet{Puglisi2019} noted that fraction of main sequence galaxies exhibiting compact star forming regions increases significantly for the most massive cases of their sample ($\mstar \gtrsim 10^{11}\msun$). Similar results are reported on the EC class of \citet[i.e., compact gas in extended stellar component]{Lyu2025}, supporting the idea of a mass-dependence of the compaction-driven evolution of star forming galaxies. The authors interpreted these galaxies as early post-starbursts, therefore possibly connected to the SBMSs (recall \sect{stats}).

\subsection{Dependence on the sub-grid recipe of star formation}
\label{sec:subgrid}

The sub-grid model of star formation implemented in \nh assumes that the unresolved gas content of a cell can be described by an empirical model calibrated on closed-box simulations of the ISM \citep[see][]{Dubois2021}. Doing so requires choosing several properties of the unresolved gas, mainly its density distribution and turbulence spectrum. In the model used here, the unresolved density PDF is assumed to be log-normal, with a variance set by the Mach number and with energy equipartition between the solenoidal and compressive modes of turbulence \citep{Federrath2012}. The density threshold for star formation in this unresolved medium is set by the product of the Mach number and the virial parameter, computed as the ratio of thermal and turbulent pressures to gas self-gravity. These choices then set the star formation efficiency of each gas cell (denser than a threshold of $10\cc$, arbitrarily chosen).

By assuming that closed-box ISM models can be directly scaled to represent the unresolved content of a cell in cosmological or galactic simulations, one necessarily underestimates the coupling between these scales. For instance, shear and tides --which generally stabilize the gas-- and the external potential from stars and dark matter --which has the opposite effect-- are not accounted for in our sub-grid model. Furthermore, the assumption of a log-normal shape with equipartition of turbulent modes for the unresolved density PDF ignores possible variations caused by compression and external forcing of turbulence. Such assumption would then lead to mis-respresentations of the true, unresolved distribution of gas over timescales depending on the nature of the forcing and the associated decay of turbulence. This is particularly critical in compressive environments (mergers, galactic centers, spiral arms) and also varies with redshift \citep{Renaud2019b, Renaud2022}. The consequence is a minimization of the diversity of turbulence spectra, and thus a tendency toward the homogenization of star formation regimes.

Therefore, the adopted sub-grid recipe may artificially reduce the importance of non-standard star formation regimes, such as starbursts, notably in terms of the intensity of the bursts. Predicting how our quantitative conclusions are affected likely requires a full understanding of the physics of scale coupling between the simulation resolution ($34 \pc$) and the scale of star formation itself. For instance, the nature of the turbulent cascade from kpc-scale injection to dissipation, and the role of external forcing, remain to be understood. Higher-resolution simulations that capture smaller scales (at the cost of sacrificing simulation volume and thus environmental diversity), as well as observations that (marginally) resolve star-forming regions, report wide ranges for the relevant quantities such as the star formation efficiency per free-fall time, the virial parameter, the Mach number, and the index of the turbulence power spectrum \citep{Renaud2013b, Lee2016, Grisdale2019, Rosolowsky2021, Segovia2025}.

Acting as a regulator of star formation, stellar feedback is expected to reduce the impact of these artefacts on star formation activity. However, the delay between star formation and feedback can be comparable to the dynamical timescales of star-forming regions, especially in fast-evolving environments such as mergers. The structure and properties of the ISM may then differ significantly between the onset of star formation and its expected regulation. In addition, uncertainties persist regarding the efficiency and scale of feedback coupling with the ISM \citep{Agertz2015, Naab2017, Grisdale2017, Ohlin2019}, again notably in dense environments such as mergers and in dense, clumpy galaxies at cosmic noon \citep{Beraldo2020, Rathjen2023}. Sub-grid recipes for stellar feedback, often calibrated on detailed observations of the Solar neighborhood, may not be equally adequate for a wider range of physical conditions (density, pressure, porosity, turbulence), and thus for a wider diversity of star formation regimes. \citet{Martin2025} noted significant discrepancies between the structures and star formation histories of galaxies in \nh and Illustris TNG50 \citep{Pillepich2019}, and argued that differences in feedback implementations could explain these variations \citep[see also][on the {\sc Firebox} simulation]{Mercado2025}. In particular, pre-supernova feedback (missing in \nh), and small-scale coupling with the ISM (absent in TNG50) are known to play an important role in the structure of the ISM and the regulation of gas consumption. Other ingredients of feedback, including cosmic rays, have also been shown to influence the large-scale galactic properties \citep{Martin2023, Bieri2026}.

In conclusion, employing sub-grid models for star formation and feedback that do not fully account for the diversity of physical conditions and environments necessarily leads to a homogenization of the resulting star formation activity. However, the existence of various modes of star formation is captured in the simulation and in our analysis, including variations across cosmic time \citep[see also][in the context of the evolution of the Kennicutt–Schmidt relation]{Kraljic2024}. This suggests that the qualitative scenario we propose is robust, although the timescales and stochasticity of these modes may vary when adopting a different recipe. Improving current models toward a framework that treats the diversity of environmental conditions across galaxies and cosmic time remains a complex task, owing to the enormous range of scales involved.

\section{Conclusion}
\label{sec:conclusions}

The detection of galaxies with starburst-like properties (e.g., compact star formation and short depletion time) within the main sequence of galaxy formation raises questions on the diversity of star formation regimes and the transition between them. By analyzing the population of these SBMS galaxies in the cosmological simulation \nh, we test and complement the hypotheses derived from large observational surveys. Our main findings are as follows.

\begin{itemize}
\item The SBMS regime exists at all redshifts ($z < 6$), and the fraction of galaxies concerned remains approximately constant ($\approx 5\%$). Along its lifetime, a galaxy can experience multiple SBMS episodes, such that, at low redshift, the majority of galaxies have been through at least one SBMS phase.
\item SBMS galaxies experience more numerous and more frequent major and minor mergers, and (on average) encounter more massive companions than any other class of galaxies, possibly because of a peculiar location in the cosmic web. Their depletion time is significantly shorter than that of the other galaxies when averaging over the entire lifetime, but also during the merger-induced starburst phases only. As a result, the SBMS galaxies assemble their stellar mass early via accretion of their neighboring galaxies and merger-triggered starbursts. This precocious build-up of their stellar component favors their location within the main sequence (via a moderate sSFR).
\item Examining individual cases reveals highly stochastic evolutions caused by rapidly changing SFRs, without the possibility of identifying a representative, general evolutionary track in the \mstar-SFR plane.
\item The SBMS phase comprises a diversity of evolutionary paths. Contrary to hypotheses in the literature, the SBMS regime denotes a slowing down of the star formation activity (from starburst to main sequence, and possibly toward quenching) for at most 25\% of the galaxies. Instead, most galaxies, especially the most massive ones, return to their previous regime after a short SBMS episode ($\approx 30\Myr$).
\item The compaction of the star-forming gas phase detected in the context of SBMS episodes leads to an increase in the SFR, but not a significant decrease in the depletion time. Such events only concern galaxies transitioning from the main sequence to starbursts.
\end{itemize}

Despite being short-lived, SBMS episodes mark important transitions along the evolution of galaxies. Their role in compaction toward quenching of star formation seems however to concern only a quarter of the galaxies. Instead, our scenario underlines the importance of a precocious assembly of the stellar mass by early mergers and merger-triggered starbursts. However, our work highlights a diversity and complexity of cases, subject to rapid variations. Establishing general evolutionary tracks for galaxies in a \mstar-SFR diagram appears challenging and not representative of individual histories. We strongly suspect that the same assessment could be made when focusing on other phases of galaxy evolution, notably when considering the diversity of physical conditions leading to quenching.

\begin{acknowledgements}
We thank the referees for their insightful reports, which helped improving the clarity of this manuscript. CA acknowledges support from the Interdisciplinary Thematic Institute IRMIA++ within the framework of the French Investments for the Future Program. We acknowledge access to the HPC Infinity Cluster hosted at the Institut d’Astrophysique de Paris and managed by S. Rouberol.
\end{acknowledgements}

\bibliographystyle{aa}
\bibliography{biblio}

@article{Song2026,
  author        = {{Song}, Jie and {Wang}, Enci and {Jia}, Cheng and {Lyu}, Cheqiu and {Chen}, Yangyao and {Wang}, Jinyang and {Li}, Fujia and {Ding}, Weiyu and {Fang}, Guanwen and {Kong}, Xu},
  title         = {{Transition from outside-in to inside-out at z {\ensuremath{\sim}} 2: Evidence from radial profiles of the specific star formation rate based on JWST/HST}},
  journal       = {\aap},
  year          = 2026,
  month         = may,
  volume        = {709},
  eid           = {A205},
  pages         = {A205},
  doi           = {10.1051/0004-6361/202658924},
  archiveprefix = {arXiv},
  eprint        = {2512.01684},
  primaryclass  = {astro-ph.GA},
  adsurl        = {https://ui.adsabs.harvard.edu/abs/2026A&A...709A.205S}
}

@article{Rinaldi2025,
  author        = {{Rinaldi}, P. and {Navarro-Carrera}, R. and {Caputi}, K.~I. and {Iani}, E. and {{\"O}stlin}, G. and {Colina}, L. and {Alberts}, S. and {{\'A}lvarez-M{\'a}rquez}, J. and {Annunziatella}, M. and {Boogaard}, L. and {Costantin}, L. and {Hjorth}, J. and {Langeroodi}, D. and {Melinder}, J. and {Moutard}, T. and {Walter}, F.},
  title         = {{The Emergence of the Star Formation Main Sequence with Redshift Unfolded by JWST}},
  journal       = {\apj},
  year          = 2025,
  month         = mar,
  volume        = {981},
  number        = {2},
  eid           = {161},
  pages         = {161},
  doi           = {10.3847/1538-4357/adb309},
  archiveprefix = {arXiv},
  eprint        = {2406.13554},
  primaryclass  = {astro-ph.GA},
  adsurl        = {https://ui.adsabs.harvard.edu/abs/2025ApJ...981..161R}
}

@article{Clarke2025,
  author        = {{Clarke}, Leonardo and {Shapley}, Alice E. and {Lam}, Natalie and {Topping}, Michael W. and {Brammer}, Gabriel B. and {Sanders}, Ryan L. and {Reddy}, Naveen A. and {Karthikeyan}, Shreya},
  title         = {{The Star-forming Main Sequence and Bursty Star-formation Histories at $z>1.4$ in JADES and AURORA}},
  journal       = {arXiv e-prints},
  year          = 2025,
  month         = oct,
  eid           = {arXiv:2510.06681},
  pages         = {arXiv:2510.06681},
  doi           = {10.48550/arXiv.2510.06681},
  archiveprefix = {arXiv},
  eprint        = {2510.06681},
  primaryclass  = {astro-ph.GA},
  adsurl        = {https://ui.adsabs.harvard.edu/abs/2025arXiv251006681C}
}

@article{Simmonds2025,
  author        = {{Simmonds}, C. and {Tacchella}, S. and {McClymont}, W. and {Curtis-Lake}, E. and {D'Eugenio}, F. and {Hainline}, K. and {Johnson}, B.~D. and {Kravtsov}, A. and {Pusk{\'a}s}, D. and {Robertson}, B. and {Stoffers}, A. and {Willott}, C. and {Baker}, W.~M. and {Belokurov}, V.~A. and {Bhatawdekar}, R. and {Bunker}, A.~J. and {Carniani}, S. and {Chevallard}, J. and {Curti}, M. and {Duan}, Q. and {Helton}, J.~M. and {Ji}, Z. and {Looser}, T.~J. and {Maiolino}, R. and {Maseda}, M.~V. and {Shivaei}, I. and {Williams}, C.~C.},
  title         = {{Bursting at the seams: the star-forming main sequence and its scatter at z = 3─9 using NIRCam photometry from JADES}},
  journal       = {\mnras},
  year          = 2025,
  month         = dec,
  volume        = {544},
  number        = {4},
  pages         = {4551-4575},
  doi           = {10.1093/mnras/staf1950},
  archiveprefix = {arXiv},
  eprint        = {2508.04410},
  primaryclass  = {astro-ph.GA},
  adsurl        = {https://ui.adsabs.harvard.edu/abs/2025MNRAS.544.4551S}
}

@article{Ellison2026,
  author        = {{Ellison}, Sara L. and {Ferreira}, Leonardo},
  title         = {{Galaxy evolution in the post-merger regime. IV - The long-term effect of mergers on galactic stellar mass growth and distribution}},
  journal       = {The Open Journal of Astrophysics},
  year          = 2026,
  month         = apr,
  volume        = {9},
  pages         = {61257},
  doi           = {10.33232/001c.161257},
  archiveprefix = {arXiv},
  eprint        = {2511.21512},
  primaryclass  = {astro-ph.GA},
  adsurl        = {https://ui.adsabs.harvard.edu/abs/2026OJAp....961257E}
}

@article{Accard2026,
  author        = {{Accard}, C{\'e}dric and {Renaud}, Florent and {Kraljic}, Katarina and {Ismail}, Diana and {B{\'e}thermin}, Matthieu and {Agertz}, Oscar},
  title         = {{Cosmic evolution of the [CII]-to-molecular gas relation}},
  journal       = {arXiv e-prints},
  year          = 2026,
  month         = apr,
  eid           = {arXiv:2604.19529},
  pages         = {arXiv:2604.19529},
  doi           = {10.48550/arXiv.2604.19529},
  archiveprefix = {arXiv},
  eprint        = {2604.19529},
  primaryclass  = {astro-ph.GA},
  adsurl        = {https://ui.adsabs.harvard.edu/abs/2026arXiv260419529A}
}

@article{Bieri2026,
  author        = {{Bieri}, Rebekka and {Pakmor}, R{\"u}diger and {van de Voort}, Freeke and {Talbot}, Rosie Y. and {Werhahn}, Maria and {Pfrommer}, Christoph and {Springel}, Volker},
  title         = {{Unveiling the Impact of Cosmic Rays on the Disc Sizes and Outflows from Dwarf Scales to Galaxy Groups}},
  journal       = {\mnras},
  year          = 2026,
  month         = feb,
  doi           = {10.1093/mnras/stag216},
  archiveprefix = {arXiv},
  eprint        = {2509.07124},
  primaryclass  = {astro-ph.GA},
  adsurl        = {https://ui.adsabs.harvard.edu/abs/2026MNRAS.tmp..214B}
}

@article{Martin2023,
  author        = {{Martin-Alvarez}, Sergio and {Sijacki}, Debora and {Haehnelt}, Martin G. and {Farcy}, Marion and {Dubois}, Yohan and {Belokurov}, Vasily and {Rosdahl}, Joakim and {Lopez-Rodriguez}, Enrique},
  title         = {{The Pandora project - I. The impact of radiation, magnetic fields, and cosmic rays on the baryonic and dark matter properties of dwarf galaxies}},
  journal       = {\mnras},
  year          = 2023,
  month         = nov,
  volume        = {525},
  number        = {3},
  pages         = {3806-3830},
  doi           = {10.1093/mnras/stad2559},
  archiveprefix = {arXiv},
  eprint        = {2211.09139},
  primaryclass  = {astro-ph.GA},
  adsurl        = {https://ui.adsabs.harvard.edu/abs/2023MNRAS.525.3806M}
}

@article{Mercado2025,
  author        = {{Mercado}, Francisco J. and {Moreno}, Jorge and {Feldmann}, Robert and {Zeender}, Marckie and {Benavides}, Jos{\'e} A. and {Piotrowska}, Joanna M. and {Klein}, Courtney and {Wheeler}, Coral and {Necib}, Lina and {Bullock}, James S. and {Hopkins}, Philip F.},
  title         = {{Effects of Galactic Environment on Size and Dark Matter Content in Low-mass Galaxies}},
  journal       = {\apj},
  year          = 2025,
  month         = apr,
  volume        = {983},
  number        = {2},
  eid           = {93},
  pages         = {93},
  doi           = {10.3847/1538-4357/adbf07},
  archiveprefix = {arXiv},
  eprint        = {2501.04084},
  primaryclass  = {astro-ph.GA},
  adsurl        = {https://ui.adsabs.harvard.edu/abs/2025ApJ...983...93M}
}

@article{Dubois2014b,
  author        = {{Dubois}, Yohan and {Volonteri}, Marta and {Silk}, Joseph and {Devriendt}, Julien and {Slyz}, Adrianne},
  title         = {{Black hole evolution - II. Spinning black holes in a supernova-driven turbulent interstellar medium}},
  journal       = {\mnras},
  year          = 2014,
  month         = may,
  volume        = {440},
  number        = {3},
  pages         = {2333-2346},
  doi           = {10.1093/mnras/stu425},
  archiveprefix = {arXiv},
  eprint        = {1401.1220},
  primaryclass  = {astro-ph.GA},
  adsurl        = {https://ui.adsabs.harvard.edu/abs/2014MNRAS.440.2333D}
}

@article{Tan2024,
  author        = {{Tan}, Qing-Hua and {Daddi}, Emanuele and {Magnelli}, Benjamin and {Correa}, Camila A. and {Bournaud}, Fr{\'e}d{\'e}ric and {Adscheid}, Sylvia and {Zhang}, Shao-Bo and {Elbaz}, David and {G{\'o}mez-Guijarro}, Carlos and {Kalita}, Boris S. and {Liu}, Daizhong and {Liu}, Zhaoxuan and {Pety}, J{\'e}r{\^o}me and {Puglisi}, Annagrazia and {Schinnerer}, Eva and {Silverman}, John D. and {Valentino}, Francesco},
  title         = {{In situ spheroid formation in distant submillimetre-bright galaxies}},
  journal       = {\nat},
  year          = 2024,
  month         = dec,
  volume        = {636},
  number        = {8041},
  pages         = {69-74},
  doi           = {10.1038/s41586-024-08201-6},
  archiveprefix = {arXiv},
  eprint        = {2407.16578},
  primaryclass  = {astro-ph.GA},
  adsurl        = {https://ui.adsabs.harvard.edu/abs/2024Natur.636...69T}
}

@article{Lebail2024,
  author        = {{Le Bail}, Aur{\'e}lien and {Daddi}, Emanuele and {Elbaz}, David and {Dickinson}, Mark and {Giavalisco}, Mauro and {Magnelli}, Benjamin and {G{\'o}mez-Guijarro}, Carlos and {Kalita}, Boris S. and {Koekemoer}, Anton M. and {Holwerda}, Benne W. and {Bournaud}, Fr{\'e}d{\'e}ric and {de la Vega}, Alexander and {Calabr{\`o}}, Antonello and {Dekel}, Avishai and {Cheng}, Yingjie and {Bisigello}, Laura and {Franco}, Maximilien and {Costantin}, Luca and {Lucas}, Ray A. and {P{\'e}rez-Gonz{\'a}lez}, Pablo G. and {Lu}, Shiying and {Wilkins}, Stephen M. and {Arrabal Haro}, Pablo and {Bagley}, Micaela B. and {Finkelstein}, Steven L. and {Kartaltepe}, Jeyhan S. and {Papovich}, Casey and {Pirzkal}, Nor and {Yung}, L.~Y. Aaron},
  title         = {{JWST/CEERS sheds light on dusty star-forming galaxies: Forming bulges, lopsidedness, and outside-in quenching at cosmic noon}},
  journal       = {\aap},
  year          = 2024,
  month         = aug,
  volume        = {688},
  eid           = {A53},
  pages         = {A53},
  doi           = {10.1051/0004-6361/202347465},
  archiveprefix = {arXiv},
  eprint        = {2307.07599},
  primaryclass  = {astro-ph.GA},
  adsurl        = {https://ui.adsabs.harvard.edu/abs/2024A&A...688A..53L}
}

@article{Daddi2022,
  author        = {{Daddi}, E. and {Delvecchio}, I. and {Dimauro}, P. and {Magnelli}, B. and {Gomez-Guijarro}, C. and {Coogan}, R. and {Elbaz}, D. and {Kalita}, B.~S. and {Le Bail}, A. and {Rich}, R.~M. and {Tan}, Q.},
  title         = {{The bending of the star-forming main sequence traces the cold- to hot-accretion transition mass over 0 < z < 4}},
  journal       = {\aap},
  year          = 2022,
  month         = may,
  volume        = {661},
  eid           = {L7},
  pages         = {L7},
  doi           = {10.1051/0004-6361/202243574},
  archiveprefix = {arXiv},
  eprint        = {2203.10880},
  primaryclass  = {astro-ph.CO},
  adsurl        = {https://ui.adsabs.harvard.edu/abs/2022A&A...661L...7D}
}

@article{Liu2025,
  author        = {{Liu}, Zhaoxuan and {Silverman}, John D. and {Daddi}, Emanuele and {Kalita}, Boris S. and {Puglisi}, Annagrazia and {Fei}, Qinyue and {Renzini}, Alvio and {Kashino}, Daichi and {Valentino}, Francesco and {Kartaltepe}, Jeyhan S. and {Liu}, Daizhong and {P{\'e}rez-Gonz{\'a}lez}, Pablo G. and {McKinney}, Jed and {Casey}, Caitlin M. and {Ding}, Xuheng and {Faisst}, Andreas and {Franco}, Maximilien and {Kakkad}, Darshan and {Koekemoer}, Anton M. and {Lambrides}, Erini and {Gillman}, Steven and {Gozaliasl}, Ghassem and {McCracken}, Henry Joy and {Rhodes}, Jason and {Robertson}, Brant E. and {Rodighiero}, Giulia and {Rujopakarn}, Wiphu and {Suzuki}, Tomoko L. and {Tanaka}, Takumi S. and {Vanderhoof}, Brittany N. and {Vijayan}, Aswin P. and {Cooper}, Olivia R. and {Kaminsky}, Aidan and {Magdis}, Georgios E. and {Roy}, Namrata},
  title         = {{A PAH deficit in the starburst core of a distant spiral galaxy}},
  journal       = {\mnras},
  year          = 2025,
  month         = sep,
  volume        = {542},
  number        = {1},
  pages         = {397-408},
  doi           = {10.1093/mnras/staf1248},
  archiveprefix = {arXiv},
  eprint        = {2505.09728},
  primaryclass  = {astro-ph.GA},
  adsurl        = {https://ui.adsabs.harvard.edu/abs/2025MNRAS.542..397L}
}

@article{Mancini2019,
  author        = {{Mancini}, Chiara and {Daddi}, Emanuele and {Juneau}, St{\'e}phanie and {Renzini}, Alvio and {Rodighiero}, Giulia and {Cappellari}, Michele and {Rodr{\'\i}guez-Mu{\~n}oz}, Luc{\'\i}a and {Liu}, Daizhong and {Pannella}, Maurilio and {Baronchelli}, Ivano and {Franceschini}, Alberto and {Bergamini}, Pietro and {D'Eugenio}, Chiara and {Puglisi}, Annagrazia},
  title         = {{Rejuvenated galaxies with very old bulges at the origin of the bending of the main sequence and of the `green valley'}},
  journal       = {\mnras},
  year          = 2019,
  month         = oct,
  volume        = {489},
  number        = {1},
  pages         = {1265-1290},
  doi           = {10.1093/mnras/stz2130},
  archiveprefix = {arXiv},
  eprint        = {1901.04573},
  primaryclass  = {astro-ph.GA},
  adsurl        = {https://ui.adsabs.harvard.edu/abs/2019MNRAS.489.1265M}
}

@article{Whitaker2014,
  author        = {{Whitaker}, Katherine E. and {Franx}, Marijn and {Leja}, Joel and {van Dokkum}, Pieter G. and {Henry}, Alaina and {Skelton}, Rosalind E. and {Fumagalli}, Mattia and {Momcheva}, Ivelina G. and {Brammer}, Gabriel B. and {Labb{\'e}}, Ivo and {Nelson}, Erica J. and {Rigby}, Jane R.},
  title         = {{Constraining the Low-mass Slope of the Star Formation Sequence at 0.5 < z < 2.5}},
  journal       = {\apj},
  year          = 2014,
  month         = nov,
  volume        = {795},
  number        = {2},
  eid           = {104},
  pages         = {104},
  doi           = {10.1088/0004-637X/795/2/104},
  archiveprefix = {arXiv},
  eprint        = {1407.1843},
  primaryclass  = {astro-ph.GA},
  adsurl        = {https://ui.adsabs.harvard.edu/abs/2014ApJ...795..104W}
}

@article{Elbaz2011,
  author        = {{Elbaz}, D. and {Dickinson}, M. and {Hwang}, H.~S. and {D{\'\i}az-Santos}, T. and {Magdis}, G. and {Magnelli}, B. and {Le Borgne}, D. and {Galliano}, F. and {Pannella}, M. and {Chanial}, P. and {Armus}, L. and {Charmandaris}, V. and {Daddi}, E. and {Aussel}, H. and {Popesso}, P. and {Kartaltepe}, J. and {Altieri}, B. and {Valtchanov}, I. and {Coia}, D. and {Dannerbauer}, H. and {Dasyra}, K. and {Leiton}, R. and {Mazzarella}, J. and {Alexander}, D.~M. and {Buat}, V. and {Burgarella}, D. and {Chary}, R.-R. and {Gilli}, R. and {Ivison}, R.~J. and {Juneau}, S. and {Le Floc'h}, E. and {Lutz}, D. and {Morrison}, G.~E. and {Mullaney}, J.~R. and {Murphy}, E. and {Pope}, A. and {Scott}, D. and {Brodwin}, M. and {Calzetti}, D. and {Cesarsky}, C. and {Charlot}, S. and {Dole}, H. and {Eisenhardt}, P. and {Ferguson}, H.~C. and {F{\"o}rster Schreiber}, N. and {Frayer}, D. and {Giavalisco}, M. and {Huynh}, M. and {Koekemoer}, A.~M. and {Papovich}, C. and {Reddy}, N. and {Surace}, C. and {Teplitz}, H. and {Yun}, M.~S. and {Wilson}, G.},
  title         = {{GOODS-Herschel: an infrared main sequence for star-forming galaxies}},
  journal       = {\aap},
  year          = 2011,
  month         = sep,
  volume        = {533},
  eid           = {A119},
  pages         = {A119},
  doi           = {10.1051/0004-6361/201117239},
  archiveprefix = {arXiv},
  eprint        = {1105.2537},
  primaryclass  = {astro-ph.CO},
  adsurl        = {https://ui.adsabs.harvard.edu/abs/2011A&A...533A.119E}
}

@article{Gui2025,
  author        = {{Gui}, Yuqian and {Xu}, Dandan and {Wang}, Haoyi and {Mei}, Xuelun and {Wang}, Enci and {Li}, Cheng and {Wuyts}, Stijn},
  title         = {{Episodic Star Formation -- I. Overview and Scatter of the Star-Forming Main Sequence}},
  journal       = {arXiv e-prints},
  year          = 2025,
  month         = nov,
  eid           = {arXiv:2512.00151},
  pages         = {arXiv:2512.00151},
  doi           = {10.48550/arXiv.2512.00151},
  archiveprefix = {arXiv},
  eprint        = {2512.00151},
  primaryclass  = {astro-ph.GA},
  adsurl        = {https://ui.adsabs.harvard.edu/abs/2025arXiv251200151G}
}

@article{Martin2025,
  author        = {{Martin}, G. and {Watkins}, A.~E. and {Dubois}, Y. and {Devriendt}, J. and {Kaviraj}, S. and {Kim}, D. and {Kraljic}, K. and {Lazar}, I. and {Pearce}, F.~R. and {Peirani}, S. and {Pichon}, C. and {Slyz}, A. and {Yi}, S.~K.},
  title         = {{Cosmic reflections I: the structural diversity of simulated and observed low-mass galaxy analogues}},
  journal       = {\mnras},
  year          = 2025,
  month         = aug,
  volume        = {541},
  number        = {2},
  pages         = {1831-1850},
  doi           = {10.1093/mnras/staf1092},
  archiveprefix = {arXiv},
  eprint        = {2505.04509},
  primaryclass  = {astro-ph.GA},
  adsurl        = {https://ui.adsabs.harvard.edu/abs/2025MNRAS.541.1831M}
}

@article{Accard2025,
  author        = {{Accard}, C. and {B{\'e}thermin}, M. and {Boquien}, M. and {Buat}, V. and {Vallini}, L. and {Renaud}, F. and {Kraljic}, K. and {Aravena}, M. and {Cassata}, P. and {da Cunha}, E. and {Dam}, P. and {de Looze}, I. and {Dessauges-Zavadsky}, M. and {Dubois}, Y. and {Faisst}, A. and {Fudamoto}, Y. and {Ginolfi}, M. and {Gruppioni}, C. and {Han}, S. and {Herrera-Camus}, R. and {Inami}, H. and {Koekemoer}, A.~M. and {Lemaux}, B.~C. and {Li}, J. and {Li}, Y. and {Mobasher}, B. and {Molina}, J. and {Nanni}, A. and {Palla}, M. and {Pozzi}, F. and {Rela{\~n}o}, M. and {Romano}, M. and {Sawant}, P. and {Spilker}, J. and {Tsujita}, A. and {Veraldi}, E. and {Villanueva}, V. and {Wang}, W. and {Yi}, S.~K. and {Zamorani}, G.},
  title         = {{The ALPINE-CRISTAL-JWST survey: Spatially resolved star formation relations at z {\ensuremath{\sim}} 5}},
  journal       = {\aap},
  year          = 2025,
  month         = oct,
  volume        = {702},
  eid           = {A206},
  pages         = {A206},
  doi           = {10.1051/0004-6361/202556140},
  archiveprefix = {arXiv},
  eprint        = {2508.13136},
  primaryclass  = {astro-ph.GA},
  adsurl        = {https://ui.adsabs.harvard.edu/abs/2025A&A...702A.206A}
}

@article{Tweed2009,
  author        = {{Tweed}, D. and {Devriendt}, J. and {Blaizot}, J. and {Colombi}, S. and {Slyz}, A.},
  title         = {{Building merger trees from cosmological N-body simulations. Towards improving galaxy formation models using subhaloes}},
  journal       = {\aap},
  year          = 2009,
  month         = nov,
  volume        = {506},
  number        = {2},
  pages         = {647-660},
  doi           = {10.1051/0004-6361/200911787},
  archiveprefix = {arXiv},
  eprint        = {0902.0679},
  primaryclass  = {astro-ph.CO},
  adsurl        = {https://ui.adsabs.harvard.edu/abs/2009A&A...506..647T}
}

@article{Tarrasse2025,
  author        = {{Tarrasse}, Maxime and {G{\'o}mez-Guijarro}, Carlos and {Elbaz}, David and {Magnelli}, Benjamin and {Dickinson}, Mark and {Henry}, Aur{\'e}lien and {Franco}, Maximilien and {Lyu}, Yipeng and {Billand}, Jean-Baptiste and {Bhatawdekar}, Rachana and {Cheng}, Yingjie and {Fontana}, Adriano and {Finkelstein}, Steven L. and {Gandolfi}, Giovanni and {Hathi}, Nimish and {Hirschmann}, Michaela and {Holwerda}, Benne W. and {Koekemoer}, Anton M. and {Lucas}, Ray A. and {Seill{\'e}}, Lise-Marie and {Wilkins}, Stephen and {Yung}, L.~Y. Aaron},
  title         = {{Compact dust-obscured star formation and the origin of the galaxy bimodality}},
  journal       = {\aap},
  year          = 2025,
  month         = may,
  volume        = {697},
  eid           = {A181},
  pages         = {A181},
  doi           = {10.1051/0004-6361/202452869},
  archiveprefix = {arXiv},
  eprint        = {2411.00279},
  primaryclass  = {astro-ph.GA},
  adsurl        = {https://ui.adsabs.harvard.edu/abs/2025A&A...697A.181T}
}

@article{Ellison2025,
  author        = {{Ellison}, Sara and {Ferreira}, Leonardo and {Bickley}, Robert and {Grindlay}, Tess and {Salim}, Samir and {Byrne-Mamahit}, Shoshannah and {Satyapal}, Shobita and {Patton}, David R. and {Scudder}, Jillian M.},
  title         = {{Galaxy evolution in the post-merger regime. III {\textendash} The triggering of active galactic nuclei peaks immediately after coalescence}},
  journal       = {The Open Journal of Astrophysics},
  year          = 2025,
  month         = feb,
  volume        = {8},
  eid           = {12},
  pages         = {12},
  doi           = {10.33232/001c.129235},
  archiveprefix = {arXiv},
  eprint        = {2412.02804},
  primaryclass  = {astro-ph.GA},
  adsurl        = {https://ui.adsabs.harvard.edu/abs/2025OJAp....8E..12E}
}

@article{Tsuge2020,
  author        = {{Tsuge}, Kisetsu and {Sano}, Hidetoshi and {Tachihara}, Kengo and {Bekki}, Kenji and {Tokuda}, Kazuki and {Inoue}, Tsuyoshi and {Mizuno}, Norikazu and {Kawamura}, Akiko and {Onishi}, Toshikazu and {Fukui}, Yasuo},
  title         = {{Active star formation across the whole Large Magellanic Cloud triggered by tidally-driven colliding HI flows}},
  journal       = {arXiv e-prints},
  year          = 2020,
  month         = oct,
  eid           = {arXiv:2010.08816},
  pages         = {arXiv:2010.08816},
  doi           = {10.48550/arXiv.2010.08816},
  archiveprefix = {arXiv},
  eprint        = {2010.08816},
  primaryclass  = {astro-ph.GA},
  adsurl        = {https://ui.adsabs.harvard.edu/abs/2020arXiv201008816T}
}

@article{Tacchella2020,
  author        = {{Tacchella}, Sandro and {Forbes}, John C. and {Caplar}, Neven},
  title         = {{Stochastic modelling of star-formation histories II: star-formation variability from molecular clouds and gas inflow}},
  journal       = {\mnras},
  year          = 2020,
  month         = sep,
  volume        = {497},
  number        = {1},
  pages         = {698-725},
  doi           = {10.1093/mnras/staa1838},
  archiveprefix = {arXiv},
  eprint        = {2006.09382},
  primaryclass  = {astro-ph.GA},
  adsurl        = {https://ui.adsabs.harvard.edu/abs/2020MNRAS.497..698T}
}

@article{Leja2019,
  author        = {{Leja}, Joel and {Carnall}, Adam C. and {Johnson}, Benjamin D. and {Conroy}, Charlie and {Speagle}, Joshua S.},
  title         = {{How to Measure Galaxy Star Formation Histories. II. Nonparametric Models}},
  journal       = {\apj},
  year          = 2019,
  month         = may,
  volume        = {876},
  number        = {1},
  eid           = {3},
  pages         = {3},
  doi           = {10.3847/1538-4357/ab133c},
  archiveprefix = {arXiv},
  eprint        = {1811.03637},
  primaryclass  = {astro-ph.GA},
  adsurl        = {https://ui.adsabs.harvard.edu/abs/2019ApJ...876....3L}
}

@article{Ocvirk2006,
  author        = {{Ocvirk}, P. and {Pichon}, C. and {Lan{\c{c}}on}, A. and {Thi{\'e}baut}, E.},
  title         = {{STECKMAP: STEllar Content and Kinematics from high resolution galactic spectra via Maximum A Posteriori}},
  journal       = {\mnras},
  year          = 2006,
  month         = jan,
  volume        = {365},
  number        = {1},
  pages         = {74-84},
  doi           = {10.1111/j.1365-2966.2005.09323.x},
  archiveprefix = {arXiv},
  eprint        = {astro-ph/0507002},
  primaryclass  = {astro-ph},
  adsurl        = {https://ui.adsabs.harvard.edu/abs/2006MNRAS.365...74O}
}

@article{Mihos1994,
  author   = {{Mihos}, J. Christopher and {Hernquist}, Lars},
  title    = {{Ultraluminous Starbursts in Major Mergers}},
  journal  = {\apjl},
  year     = 1994,
  month    = aug,
  volume   = {431},
  pages    = {L9},
  doi      = {10.1086/187460},
  adsurl   = {https://ui.adsabs.harvard.edu/abs/1994ApJ...431L...9M}
}

@article{Pan2019,
  author        = {{Pan}, Hsi-An and {Lin}, Lihwai and {Hsieh}, Bau-Ching and {Barrera-Ballesteros}, Jorge K. and {S{\'a}nchez}, Sebasti{\'a}n F. and {Hsu}, Chin-Hao and {Keenan}, Ryan and {Tissera}, Patricia B. and {Boquien}, M{\'e}d{\'e}ric and {Dai}, Y. Sophia and {Knapen}, Johan H. and {Riffel}, Rog{\'e}rio and {Argudo-Fern{\'a}ndez}, Maria and {Xiao}, Ting and {Yuan}, Fang-Ting},
  title         = {{SDSS-IV MaNGA: Spatial Evolution of Star Formation Triggered by Galaxy Interactions}},
  journal       = {\apj},
  year          = 2019,
  month         = aug,
  volume        = {881},
  number        = {2},
  eid           = {119},
  pages         = {119},
  doi           = {10.3847/1538-4357/ab311c},
  archiveprefix = {arXiv},
  eprint        = {1907.04491},
  primaryclass  = {astro-ph.GA},
  adsurl        = {https://ui.adsabs.harvard.edu/abs/2019ApJ...881..119P}
}

@article{Tacconi2020,
  author        = {{Tacconi}, Linda J. and {Genzel}, Reinhard and {Sternberg}, Amiel},
  title         = {{The Evolution of the Star-Forming Interstellar Medium Across Cosmic Time}},
  journal       = {\araa},
  year          = 2020,
  month         = aug,
  volume        = {58},
  pages         = {157-203},
  doi           = {10.1146/annurev-astro-082812-141034},
  archiveprefix = {arXiv},
  eprint        = {2003.06245},
  primaryclass  = {astro-ph.GA},
  adsurl        = {https://ui.adsabs.harvard.edu/abs/2020ARA&A..58..157T}
}

@article{Fabian2012,
  author        = {{Fabian}, A.~C.},
  title         = {{Observational Evidence of Active Galactic Nuclei Feedback}},
  journal       = {\araa},
  year          = 2012,
  month         = sep,
  volume        = {50},
  pages         = {455-489},
  doi           = {10.1146/annurev-astro-081811-125521},
  archiveprefix = {arXiv},
  eprint        = {1204.4114},
  primaryclass  = {astro-ph.CO},
  adsurl        = {https://ui.adsabs.harvard.edu/abs/2012ARA&A..50..455F}
}

@article{Peng2010,
  author        = {{Peng}, Ying-jie and {Lilly}, Simon J. and {Kova{\v{c}}}, Katarina and {Bolzonella}, Micol and {Pozzetti}, Lucia and {Renzini}, Alvio and {Zamorani}, Gianni and {Ilbert}, Olivier and {Knobel}, Christian and {Iovino}, Angela and {Maier}, Christian and {Cucciati}, Olga and {Tasca}, Lidia and {Carollo}, C. Marcella and {Silverman}, John and {Kampczyk}, Pawel and {de Ravel}, Loic and {Sanders}, David and {Scoville}, Nicholas and {Contini}, Thierry and {Mainieri}, Vincenzo and {Scodeggio}, Marco and {Kneib}, Jean-Paul and {Le F{\`e}vre}, Olivier and {Bardelli}, Sandro and {Bongiorno}, Angela and {Caputi}, Karina and {Coppa}, Graziano and {de la Torre}, Sylvain and {Franzetti}, Paolo and {Garilli}, Bianca and {Lamareille}, Fabrice and {Le Borgne}, Jean-Francois and {Le Brun}, Vincent and {Mignoli}, Marco and {Perez Montero}, Enrique and {Pello}, Roser and {Ricciardelli}, Elena and {Tanaka}, Masayuki and {Tresse}, Laurence and {Vergani}, Daniela and {Welikala}, Niraj and {Zucca}, Elena and {Oesch}, Pascal and {Abbas}, Ummi and {Barnes}, Luke and {Bordoloi}, Rongmon and {Bottini}, Dario and {Cappi}, Alberto and {Cassata}, Paolo and {Cimatti}, Andrea and {Fumana}, Marco and {Hasinger}, Gunther and {Koekemoer}, Anton and {Leauthaud}, Alexei and {Maccagni}, Dario and {Marinoni}, Christian and {McCracken}, Henry and {Memeo}, Pierdomenico and {Meneux}, Baptiste and {Nair}, Preethi and {Porciani}, Cristiano and {Presotto}, Valentina and {Scaramella}, Roberto},
  title         = {{Mass and Environment as Drivers of Galaxy Evolution in SDSS and zCOSMOS and the Origin of the Schechter Function}},
  journal       = {\apj},
  year          = 2010,
  month         = sep,
  volume        = {721},
  number        = {1},
  pages         = {193-221},
  doi           = {10.1088/0004-637X/721/1/193},
  archiveprefix = {arXiv},
  eprint        = {1003.4747},
  primaryclass  = {astro-ph.CO},
  adsurl        = {https://ui.adsabs.harvard.edu/abs/2010ApJ...721..193P}
}

@article{Lower2020,
  author        = {{Lower}, Sidney and {Narayanan}, Desika and {Leja}, Joel and {Johnson}, Benjamin D. and {Conroy}, Charlie and {Dav{\'e}}, Romeel},
  title         = {{How Well Can We Measure the Stellar Mass of a Galaxy: The Impact of the Assumed Star Formation History Model in SED Fitting}},
  journal       = {\apj},
  year          = 2020,
  month         = nov,
  volume        = {904},
  number        = {1},
  eid           = {33},
  pages         = {33},
  doi           = {10.3847/1538-4357/abbfa7},
  archiveprefix = {arXiv},
  eprint        = {2006.03599},
  primaryclass  = {astro-ph.GA},
  adsurl        = {https://ui.adsabs.harvard.edu/abs/2020ApJ...904...33L}
}

@article{Iyer2017,
  author        = {{Iyer}, Kartheik and {Gawiser}, Eric},
  title         = {{Reconstruction of Galaxy Star Formation Histories through SED Fitting:The Dense Basis Approach}},
  journal       = {\apj},
  year          = 2017,
  month         = apr,
  volume        = {838},
  number        = {2},
  eid           = {127},
  pages         = {127},
  doi           = {10.3847/1538-4357/aa63f0},
  archiveprefix = {arXiv},
  eprint        = {1702.04371},
  primaryclass  = {astro-ph.GA},
  adsurl        = {https://ui.adsabs.harvard.edu/abs/2017ApJ...838..127I}
}

@article{Ciesla2017,
  author        = {{Ciesla}, L. and {Elbaz}, D. and {Fensch}, J.},
  title         = {{The SFR-M$_{{\ensuremath{*}}}$ main sequence archetypal star-formation history and analytical models}},
  journal       = {\aap},
  year          = 2017,
  month         = dec,
  volume        = {608},
  eid           = {A41},
  pages         = {A41},
  doi           = {10.1051/0004-6361/201731036},
  archiveprefix = {arXiv},
  eprint        = {1706.08531},
  primaryclass  = {astro-ph.GA},
  adsurl        = {https://ui.adsabs.harvard.edu/abs/2017A&A...608A..41C}
}

@article{Conroy2013,
  author        = {{Conroy}, Charlie},
  title         = {{Modeling the Panchromatic Spectral Energy Distributions of Galaxies}},
  journal       = {\araa},
  year          = 2013,
  month         = aug,
  volume        = {51},
  number        = {1},
  pages         = {393-455},
  doi           = {10.1146/annurev-astro-082812-141017},
  archiveprefix = {arXiv},
  eprint        = {1301.7095},
  primaryclass  = {astro-ph.CO},
  adsurl        = {https://ui.adsabs.harvard.edu/abs/2013ARA&A..51..393C}
}

@article{Rathjen2023,
  author        = {{Rathjen}, Tim-Eric and {Naab}, Thorsten and {Walch}, Stefanie and {Seifried}, Daniel and {Girichidis}, Philipp and {W{\"u}nsch}, Richard},
  title         = {{SILCC - VII. Gas kinematics and multiphase outflows of the simulated ISM at high gas surface densities}},
  journal       = {\mnras},
  year          = 2023,
  month         = jun,
  volume        = {522},
  number        = {2},
  pages         = {1843-1862},
  doi           = {10.1093/mnras/stad1104},
  archiveprefix = {arXiv},
  eprint        = {2211.15419},
  primaryclass  = {astro-ph.GA},
  adsurl        = {https://ui.adsabs.harvard.edu/abs/2023MNRAS.522.1843R}
}

@article{OLeary2021,
  author        = {{O'Leary}, Joseph A. and {Moster}, Benjamin P. and {Naab}, Thorsten and {Somerville}, Rachel S.},
  title         = {{EMERGE: empirical predictions of galaxy merger rates since z {\ensuremath{\sim}} 6}},
  journal       = {\mnras},
  year          = 2021,
  month         = mar,
  volume        = {501},
  number        = {3},
  pages         = {3215-3237},
  doi           = {10.1093/mnras/staa3746},
  archiveprefix = {arXiv},
  eprint        = {2001.02687},
  primaryclass  = {astro-ph.GA},
  adsurl        = {https://ui.adsabs.harvard.edu/abs/2021MNRAS.501.3215O}
}

@article{Rosolowsky2021,
  author        = {{Rosolowsky}, Erik and {Hughes}, Annie and {Leroy}, Adam K. and {Sun}, Jiayi and {Querejeta}, Miguel and {Schruba}, Andreas and {Usero}, Antonio and {Herrera}, Cinthya N. and {Liu}, Daizhong and {Pety}, J{\'e}r{\^o}me and {Saito}, Toshiki and {Be{\v{s}}li{\'c}}, Ivana and {Bigiel}, Frank and {Blanc}, Guillermo and {Chevance}, M{\'e}lanie and {Dale}, Daniel A. and {Deger}, Sinan and {Faesi}, Christopher M. and {Glover}, Simon C.~O. and {Henshaw}, Jonathan D. and {Klessen}, Ralf S. and {Kruijssen}, J.~M. Diederik and {Larson}, Kirsten and {Lee}, Janice and {Meidt}, Sharon and {Mok}, Angus and {Schinnerer}, Eva and {Thilker}, David A. and {Williams}, Thomas G.},
  title         = {{Giant molecular cloud catalogues for PHANGS-ALMA: methods and initial results}},
  journal       = {\mnras},
  year          = 2021,
  month         = mar,
  volume        = {502},
  number        = {1},
  pages         = {1218-1245},
  doi           = {10.1093/mnras/stab085},
  archiveprefix = {arXiv},
  eprint        = {2101.04697},
  primaryclass  = {astro-ph.GA},
  adsurl        = {https://ui.adsabs.harvard.edu/abs/2021MNRAS.502.1218R}
}

@article{Segovia2025,
  author        = {{Segovia Otero}, {\'A}lvaro and {Agertz}, Oscar and {Renaud}, Florent and {Kraljic}, Katarina and {Romeo}, Alessandro B. and {Semenov}, Vadim A.},
  title         = {{Cosmic evolution of the star formation efficiency in Milky Way-like galaxies}},
  journal       = {\mnras},
  year          = 2025,
  month         = apr,
  volume        = {538},
  number        = {4},
  pages         = {2646-2659},
  doi           = {10.1093/mnras/staf423},
  archiveprefix = {arXiv},
  eprint        = {2410.08266},
  primaryclass  = {astro-ph.GA},
  adsurl        = {https://ui.adsabs.harvard.edu/abs/2025MNRAS.538.2646S}
}

@article{Zhang2025,
  author        = {{Zhang}, Zhijie and {Zhang}, Xiaoxia and {Li}, Hui and {Fang}, Taotao and {Luo}, Yang and {Marinacci}, Federico and {Sales}, Laura V. and {Torrey}, Paul and {Vogelsberger}, Mark and {Yu}, Qingzheng and {Yuan}, Feng},
  title         = {{Tracing the Origins of Hot Halo Gas in Milky Way─type Galaxies with SMUGGLE}},
  journal       = {\apj},
  year          = 2025,
  month         = oct,
  volume        = {991},
  number        = {2},
  eid           = {170},
  pages         = {170},
  doi           = {10.3847/1538-4357/ae019f},
  archiveprefix = {arXiv},
  eprint        = {2508.21576},
  primaryclass  = {astro-ph.GA},
  adsurl        = {https://ui.adsabs.harvard.edu/abs/2025ApJ...991..170Z}
}

@article{Carvajal2025,
  author        = {{Carvajal-Bohorquez}, C. and {Ciesla}, L. and {Laporte}, N. and {Boquien}, M. and {Buat}, V. and {Ilbert}, O. and {Aufort}, G. and {Shuntov}, M. and {Witten}, C. and {Oesch}, P.~A. and {Covelo-Paz}, A.},
  title         = {{Stochastic star formation activity of galaxies within the first billion years probed by JWST}},
  journal       = {arXiv e-prints},
  year          = 2025,
  month         = jul,
  eid           = {arXiv:2507.13160},
  pages         = {arXiv:2507.13160},
  doi           = {10.48550/arXiv.2507.13160},
  archiveprefix = {arXiv},
  eprint        = {2507.13160},
  primaryclass  = {astro-ph.GA},
  adsurl        = {https://ui.adsabs.harvard.edu/abs/2025arXiv250713160C}
}

@article{Caplar2019,
  author        = {{Caplar}, Neven and {Tacchella}, Sandro},
  title         = {{Stochastic modelling of star-formation histories I: the scatter of the star-forming main sequence}},
  journal       = {\mnras},
  year          = 2019,
  month         = aug,
  volume        = {487},
  number        = {3},
  pages         = {3845-3869},
  doi           = {10.1093/mnras/stz1449},
  archiveprefix = {arXiv},
  eprint        = {1901.07556},
  primaryclass  = {astro-ph.GA},
  adsurl        = {https://ui.adsabs.harvard.edu/abs/2019MNRAS.487.3845C}
}

@article{Arango2025,
  author        = {{Arango-Toro}, R.~C. and {Ilbert}, O. and {Ciesla}, L. and {Shuntov}, M. and {Aufort}, G. and {Mercier}, W. and {Laigle}, C. and {Franco}, M. and {Bethermin}, M. and {Le Borgne}, D. and {Dubois}, Y. and {McCracken}, H.~J. and {Paquereau}, L. and {Huertas-Company}, M. and {Kartaltepe}, J. and {Casey}, C.~M. and {Akins}, H. and {Allen}, N. and {Andika}, I. and {Brinch}, M. and {Drakos}, N.~E. and {Faisst}, A. and {Gozaliasl}, G. and {Harish}, S. and {Kaminsky}, A. and {Koekemoer}, A. and {Kokorev}, V. and {Liu}, D. and {Magdis}, G. and {Martin}, C.~L. and {Moutard}, T. and {Rhodes}, J. and {Rich}, R.~M. and {Robertson}, B. and {Sanders}, D.~B. and {Sheth}, K. and {Talia}, M. and {Toft}, S. and {Tresse}, L. and {Valentino}, F. and {Vijayan}, A. and {Weaver}, J.},
  title         = {{COSMOS-Web: A history of galaxy migrations over the stellar mass{\textendash}star formation rate plane}},
  journal       = {\aap},
  year          = 2025,
  month         = apr,
  volume        = {696},
  eid           = {A159},
  pages         = {A159},
  doi           = {10.1051/0004-6361/202452519},
  archiveprefix = {arXiv},
  eprint        = {2410.05375},
  primaryclass  = {astro-ph.GA},
  adsurl        = {https://ui.adsabs.harvard.edu/abs/2025A&A...696A.159A}
}

@article{Daddi2007,
  author        = {{Daddi}, E. and {Dickinson}, M. and {Morrison}, G. and {Chary}, R. and {Cimatti}, A. and {Elbaz}, D. and {Frayer}, D. and {Renzini}, A. and {Pope}, A. and {Alexander}, D.~M. and {Bauer}, F.~E. and {Giavalisco}, M. and {Huynh}, M. and {Kurk}, J. and {Mignoli}, M.},
  title         = {{Multiwavelength Study of Massive Galaxies at z\raisebox{-0.5ex}\textasciitilde2. I. Star Formation and Galaxy Growth}},
  journal       = {\apj},
  year          = 2007,
  month         = nov,
  volume        = {670},
  number        = {1},
  pages         = {156-172},
  doi           = {10.1086/521818},
  archiveprefix = {arXiv},
  eprint        = {0705.2831},
  primaryclass  = {astro-ph},
  adsurl        = {https://ui.adsabs.harvard.edu/abs/2007ApJ...670..156D}
}

@article{Jimenez2019,
  author        = {{Jim{\'e}nez-Andrade}, E.~F. and {Magnelli}, B. and {Karim}, A. and {Zamorani}, G. and {Bondi}, M. and {Schinnerer}, E. and {Sargent}, M. and {Romano-D{\'\i}az}, E. and {Novak}, M. and {Lang}, P. and {Bertoldi}, F. and {Vardoulaki}, E. and {Toft}, S. and {Smol{\v{c}}i{\'c}}, V. and {Harrington}, K. and {Leslie}, S. and {Delhaize}, J. and {Liu}, D. and {Karoumpis}, C. and {Kartaltepe}, J. and {Koekemoer}, A.~M.},
  title         = {{Radio continuum size evolution of star-forming galaxies over 0.35 < z < 2.25}},
  journal       = {\aap},
  year          = 2019,
  month         = may,
  volume        = {625},
  eid           = {A114},
  pages         = {A114},
  doi           = {10.1051/0004-6361/201935178},
  archiveprefix = {arXiv},
  eprint        = {1903.12217},
  primaryclass  = {astro-ph.GA},
  adsurl        = {https://ui.adsabs.harvard.edu/abs/2019A&A...625A.114J}
}

@article{Martin2021,
  author        = {{Martin}, G. and {Jackson}, R.~A. and {Kaviraj}, S. and {Choi}, H. and {Devriendt}, J.~E.~G. and {Dubois}, Y. and {Kimm}, T. and {Kraljic}, K. and {Peirani}, S. and {Pichon}, C. and {Volonteri}, M. and {Yi}, S.~K.},
  title         = {{The role of mergers and interactions in driving the evolution of dwarf galaxies over cosmic time}},
  journal       = {\mnras},
  year          = 2021,
  month         = jan,
  volume        = {500},
  number        = {4},
  pages         = {4937-4957},
  doi           = {10.1093/mnras/staa3443},
  archiveprefix = {arXiv},
  eprint        = {2007.07913},
  primaryclass  = {astro-ph.GA},
  adsurl        = {https://ui.adsabs.harvard.edu/abs/2021MNRAS.500.4937M}
}

@article{Lapiner2023,
  author        = {{Lapiner}, Sharon and {Dekel}, Avishai and {Freundlich}, Jonathan and {Ginzburg}, Omri and {Jiang}, Fangzhou and {Kretschmer}, Michael and {Tacchella}, Sandro and {Ceverino}, Daniel and {Primack}, Joel},
  title         = {{Wet compaction to a blue nugget: a critical phase in galaxy evolution}},
  journal       = {\mnras},
  year          = 2023,
  month         = jul,
  volume        = {522},
  number        = {3},
  pages         = {4515-4547},
  doi           = {10.1093/mnras/stad1263},
  archiveprefix = {arXiv},
  eprint        = {2302.12234},
  primaryclass  = {astro-ph.GA},
  adsurl        = {https://ui.adsabs.harvard.edu/abs/2023MNRAS.522.4515L}
}

@article{Dekel2014,
  author        = {{Dekel}, A. and {Burkert}, A.},
  title         = {{Wet disc contraction to galactic blue nuggets and quenching to red nuggets}},
  journal       = {\mnras},
  year          = 2014,
  month         = feb,
  volume        = {438},
  number        = {2},
  pages         = {1870-1879},
  doi           = {10.1093/mnras/stt2331},
  archiveprefix = {arXiv},
  eprint        = {1310.1074},
  primaryclass  = {astro-ph.CO},
  adsurl        = {https://ui.adsabs.harvard.edu/abs/2014MNRAS.438.1870D}
}

@article{Muratov2015,
  author        = {{Muratov}, Alexander L. and {Kere{\v{s}}}, Du{\v{s}}an and {Faucher-Gigu{\`e}re}, Claude-Andr{\'e} and {Hopkins}, Philip F. and {Quataert}, Eliot and {Murray}, Norman},
  title         = {{Gusty, gaseous flows of FIRE: galactic winds in cosmological simulations with explicit stellar feedback}},
  journal       = {\mnras},
  year          = 2015,
  month         = dec,
  volume        = {454},
  number        = {3},
  pages         = {2691-2713},
  doi           = {10.1093/mnras/stv2126},
  archiveprefix = {arXiv},
  eprint        = {1501.03155},
  primaryclass  = {astro-ph.GA},
  adsurl        = {https://ui.adsabs.harvard.edu/abs/2015MNRAS.454.2691M}
}

@article{Christensen2016,
  author        = {{Christensen}, Charlotte R. and {Dav{\'e}}, Romeel and {Governato}, Fabio and {Pontzen}, Andrew and {Brooks}, Alyson and {Munshi}, Ferah and {Quinn}, Thomas and {Wadsley}, James},
  title         = {{In-N-Out: The Gas Cycle from Dwarfs to Spiral Galaxies}},
  journal       = {\apj},
  year          = 2016,
  month         = jun,
  volume        = {824},
  number        = {1},
  eid           = {57},
  pages         = {57},
  doi           = {10.3847/0004-637X/824/1/57},
  archiveprefix = {arXiv},
  eprint        = {1508.00007},
  primaryclass  = {astro-ph.GA},
  adsurl        = {https://ui.adsabs.harvard.edu/abs/2016ApJ...824...57C}
}

@article{Barro2014,
  author        = {{Barro}, G. and {Faber}, S.~M. and {P{\'e}rez-Gonz{\'a}lez}, P.~G. and {Pacifici}, C. and {Trump}, J.~R. and {Koo}, D.~C. and {Wuyts}, S. and {Guo}, Y. and {Bell}, E. and {Dekel}, A. and {Porter}, L. and {Primack}, J. and {Ferguson}, H. and {Ashby}, M.~L.~N. and {Caputi}, K. and {Ceverino}, D. and {Croton}, D. and {Fazio}, G.~G. and {Giavalisco}, M. and {Hsu}, L. and {Kocevski}, D. and {Koekemoer}, A. and {Kurczynski}, P. and {Kollipara}, P. and {Lee}, J. and {McIntosh}, D.~H. and {McGrath}, E. and {Moody}, C. and {Somerville}, R. and {Papovich}, C. and {Salvato}, M. and {Santini}, P. and {Tal}, T. and {van der Wel}, A. and {Williams}, C.~C. and {Willner}, S.~P. and {Zolotov}, A.},
  title         = {{CANDELS+3D-HST: Compact SFGs at z \raisebox{-0.5ex}\textasciitilde 2-3, the Progenitors of the First Quiescent Galaxies}},
  journal       = {\apj},
  year          = 2014,
  month         = aug,
  volume        = {791},
  number        = {1},
  eid           = {52},
  pages         = {52},
  doi           = {10.1088/0004-637X/791/1/52},
  archiveprefix = {arXiv},
  eprint        = {1311.5559},
  primaryclass  = {astro-ph.CO},
  adsurl        = {https://ui.adsabs.harvard.edu/abs/2014ApJ...791...52B}
}

@article{Tadaki2020,
  author        = {{Tadaki}, Ken-ichi and {Belli}, Sirio and {Burkert}, Andreas and {Dekel}, Avishai and {F{\"o}rster Schreiber}, Natascha M. and {Genzel}, Reinhard and {Hayashi}, Masao and {Herrera-Camus}, Rodrigo and {Kodama}, Tadayuki and {Kohno}, Kotaro and {Koyama}, Yusei and {Lee}, Minju M. and {Lutz}, Dieter and {Mowla}, Lamiya and {Nelson}, Erica J. and {Renzini}, Alvio and {Suzuki}, Tomoko L. and {Tacconi}, Linda J. and {{\"U}bler}, Hannah and {Wisnioski}, Emily and {Wuyts}, Stijn},
  title         = {{Structural Evolution in Massive Galaxies at z {\ensuremath{\sim}} 2}},
  journal       = {\apj},
  year          = 2020,
  month         = sep,
  volume        = {901},
  number        = {1},
  eid           = {74},
  pages         = {74},
  doi           = {10.3847/1538-4357/abaf4a},
  archiveprefix = {arXiv},
  eprint        = {2009.01976},
  primaryclass  = {astro-ph.GA},
  adsurl        = {https://ui.adsabs.harvard.edu/abs/2020ApJ...901...74T}
}

@article{Franco2020,
  author        = {{Franco}, M. and {Elbaz}, D. and {Zhou}, L. and {Magnelli}, B. and {Schreiber}, C. and {Ciesla}, L. and {Dickinson}, M. and {Nagar}, N. and {Magdis}, G. and {Alexander}, D.~M. and {B{\'e}thermin}, M. and {Demarco}, R. and {Daddi}, E. and {Wang}, T. and {Mullaney}, J. and {Sargent}, M. and {Inami}, H. and {Shu}, X. and {Bournaud}, F. and {Chary}, R. and {Coogan}, R.~T. and {Ferguson}, H. and {Finkelstein}, S.~L. and {Giavalisco}, M. and {G{\'o}mez-Guijarro}, C. and {Iono}, D. and {Juneau}, S. and {Lagache}, G. and {Lin}, L. and {Motohara}, K. and {Okumura}, K. and {Pannella}, M. and {Papovich}, C. and {Pope}, A. and {Rujopakarn}, W. and {Silverman}, J. and {Xiao}, M.},
  title         = {{GOODS-ALMA: The slow downfall of star formation in z = 2-3 massive galaxies}},
  journal       = {\aap},
  year          = 2020,
  month         = nov,
  volume        = {643},
  eid           = {A30},
  pages         = {A30},
  doi           = {10.1051/0004-6361/202038312},
  archiveprefix = {arXiv},
  eprint        = {2005.03043},
  primaryclass  = {astro-ph.GA},
  adsurl        = {https://ui.adsabs.harvard.edu/abs/2020A&A...643A..30F}
}

@article{Popesso2023,
  author        = {{Popesso}, P. and {Concas}, A. and {Cresci}, G. and {Belli}, S. and {Rodighiero}, G. and {Inami}, H. and {Dickinson}, M. and {Ilbert}, O. and {Pannella}, M. and {Elbaz}, D.},
  title         = {{The main sequence of star-forming galaxies across cosmic times}},
  journal       = {\mnras},
  year          = 2023,
  month         = feb,
  volume        = {519},
  number        = {1},
  pages         = {1526-1544},
  doi           = {10.1093/mnras/stac3214},
  archiveprefix = {arXiv},
  eprint        = {2203.10487},
  primaryclass  = {astro-ph.GA},
  adsurl        = {https://ui.adsabs.harvard.edu/abs/2023MNRAS.519.1526P}
}

@ARTICLE{Casey2024,
       author = {{Casey}, Caitlin M. and {Akins}, Hollis B. and {Shuntov}, Marko and {Ilbert}, Olivier and {Paquereau}, Louise and {Franco}, Maximilien and {Hayward}, Christopher C. and {Finkelstein}, Steven L. and {Boylan-Kolchin}, Michael and {Robertson}, Brant E. and {Allen}, Natalie and {Brinch}, Malte and {Cooper}, Olivia R. and {Ding}, Xuheng and {Drakos}, Nicole E. and {Faisst}, Andreas L. and {Fujimoto}, Seiji and {Gillman}, Steven and {Harish}, Santosh and {Hirschmann}, Michaela and {Jin}, Shuowen and {Kartaltepe}, Jeyhan S. and {Koekemoer}, Anton M. and {Kokorev}, Vasily and {Liu}, Daizhong and {Long}, Arianna S. and {Magdis}, Georgios and {Maraston}, Claudia and {Martin}, Crystal L. and {McCracken}, Henry Joy and {McKinney}, Jed and {Mobasher}, Bahram and {Rhodes}, Jason and {Rich}, R. Michael and {Sanders}, David B. and {Silverman}, John D. and {Toft}, Sune and {Vijayan}, Aswin P. and {Weaver}, John R. and {Wilkins}, Stephen M. and {Yang}, Lilan and {Zavala}, Jorge A.},
        title = "{COSMOS-Web: Intrinsically Luminous z {\ensuremath{\gtrsim}} 10 Galaxy Candidates Test Early Stellar Mass Assembly}",
      journal = {\apj},
         year = 2024,
        month = apr,
       volume = {965},
       number = {1},
          eid = {98},
        pages = {98},
          doi = {10.3847/1538-4357/ad2075},
archivePrefix = {arXiv},
       eprint = {2308.10932},
 primaryClass = {astro-ph.GA},
       adsurl = {https://ui.adsabs.harvard.edu/abs/2024ApJ...965...98C}
}

@ARTICLE{Carniani2025,
       author = {{Carniani}, Stefano and {D'Eugenio}, Francesco and {Ji}, Xihan and {Parlanti}, Eleonora and {Scholtz}, Jan and {Sun}, Fengwu and {Venturi}, Giacomo and {Bakx}, Tom J.~L.~C. and {Curti}, Mirko and {Maiolino}, Roberto and {Tacchella}, Sandro and {Zavala}, Jorge A. and {Hainline}, Kevin and {Witstok}, Joris and {Johnson}, Benjamin D. and {Alberts}, Stacey and {Bunker}, Andrew J. and {Charlot}, St{\'e}phane and {Eisenstein}, Daniel J. and {Helton}, Jakob M. and {Jakobsen}, Peter and {Kumari}, Nimisha and {Robertson}, Brant and {Saxena}, Aayush and {{\"U}bler}, Hannah and {Williams}, Christina C. and {Willmer}, Christopher N.~A. and {Willott}, Chris},
        title = "{The eventful life of a luminous galaxy at z = 14: metal enrichment, feedback, and low gas fraction?}",
      journal = {\aap},
         year = 2025,
        month = apr,
       volume = {696},
          eid = {A87},
        pages = {A87},
          doi = {10.1051/0004-6361/202452451},
archivePrefix = {arXiv},
       eprint = {2409.20533},
 primaryClass = {astro-ph.GA},
       adsurl = {https://ui.adsabs.harvard.edu/abs/2025A&A...696A..87C}
}

@ARTICLE{Lyu2025,
       author = {{Lyu}, Yipeng and {Magnelli}, Benjamin and {Elbaz}, David and {P{\'e}rez-Gonz{\'a}lez}, Pablo G. and {Correa}, Camila and {Daddi}, Emanuele and {G{\'o}mez-Guijarro}, Carlos and {Dunlop}, James S. and {Grogin}, Norman A. and {Koekemoer}, Anton M. and {McLeod}, Derek J. and {Lu}, Shiying},
        title = "{PRIMER: JWST/MIRI reveals the evolution of star-forming structures in galaxies at z {\ensuremath{\leq}} 2.5}",
      journal = {\aap},
         year = 2025,
        month = jan,
       volume = {693},
          eid = {A313},
        pages = {A313},
          doi = {10.1051/0004-6361/202451067},
archivePrefix = {arXiv},
       eprint = {2406.11571},
 primaryClass = {astro-ph.GA},
       adsurl = {https://ui.adsabs.harvard.edu/abs/2025A&A...693A.313L}
}

@ARTICLE{Magnelli2023,
       author = {{Magnelli}, Benjamin and {G{\'o}mez-Guijarro}, Carlos and {Elbaz}, David and {Daddi}, Emanuele and {Papovich}, Casey and {Shen}, Lu and {Arrabal Haro}, Pablo and {Bagley}, Micaela B. and {Bell}, Eric F. and {Buat}, V{\'e}ronique and {Costantin}, Luca and {Dickinson}, Mark and {Finkelstein}, Steven L. and {Gardner}, Jonathan P. and {Jim{\'e}nez-Andrade}, Eric F. and {Kartaltepe}, Jeyhan S. and {Koekemoer}, Anton M. and {Lyu}, Yipeng and {P{\'e}rez-Gonz{\'a}lez}, Pablo G. and {Pirzkal}, Nor and {Tacchella}, Sandro and {de la Vega}, Alexander and {Wuyts}, Stijn and {Yang}, Guang and {Yung}, L.~Y. Aaron and {Zavala}, Jorge},
        title = "{CEERS: MIRI deciphers the spatial distribution of dust-obscured star formation in galaxies at 0.1 < z < 2.5}",
      journal = {\aap},
         year = 2023,
        month = oct,
       volume = {678},
          eid = {A83},
        pages = {A83},
          doi = {10.1051/0004-6361/202347052},
archivePrefix = {arXiv},
       eprint = {2305.19331},
 primaryClass = {astro-ph.GA},
       adsurl = {https://ui.adsabs.harvard.edu/abs/2023A&A...678A..83M}
}

@ARTICLE{Puglisi2021,
       author = {{Puglisi}, Annagrazia and {Daddi}, Emanuele and {Valentino}, Francesco and {Magdis}, Georgios and {Liu}, Daizhong and {Kokorev}, Vasily and {Circosta}, Chiara and {Elbaz}, David and {Bournaud}, Frederic and {Gomez-Guijarro}, Carlos and {Jin}, Shuowen and {Madden}, Suzanne and {Sargent}, Mark T. and {Swinbank}, Mark},
        title = "{Submillimetre compactness as a critical dimension to understand the main sequence of star-forming galaxies}",
      journal = {\mnras},
         year = 2021,
        month = dec,
       volume = {508},
       number = {4},
        pages = {5217-5238},
          doi = {10.1093/mnras/stab2914},
archivePrefix = {arXiv},
       eprint = {2103.12035},
 primaryClass = {astro-ph.GA},
       adsurl = {https://ui.adsabs.harvard.edu/abs/2021MNRAS.508.5217P}
}

@ARTICLE{Puglisi2019,
       author = {{Puglisi}, A. and {Daddi}, E. and {Liu}, D. and {Bournaud}, F. and {Silverman}, J.~D. and {Circosta}, C. and {Calabr{\`o}}, A. and {Aravena}, M. and {Cibinel}, A. and {Dannerbauer}, H. and {Delvecchio}, I. and {Elbaz}, D. and {Gao}, Y. and {Gobat}, R. and {Jin}, S. and {Le Floc'h}, E. and {Magdis}, G.~E. and {Mancini}, C. and {Riechers}, D.~A. and {Rodighiero}, G. and {Sargent}, M. and {Valentino}, F. and {Zanisi}, L.},
        title = "{The Main Sequence at z {\ensuremath{\sim}} 1.3 Contains a Sizable Fraction of Galaxies with Compact Star Formation Sizes: A New Population of Early Post-starbursts?}",
      journal = {\apjl},
         year = 2019,
        month = jun,
       volume = {877},
       number = {2},
          eid = {L23},
        pages = {L23},
          doi = {10.3847/2041-8213/ab1f92},
archivePrefix = {arXiv},
       eprint = {1905.02958},
 primaryClass = {astro-ph.GA},
       adsurl = {https://ui.adsabs.harvard.edu/abs/2019ApJ...877L..23P}
}

@ARTICLE{Tacchella2016,
       author = {{Tacchella}, Sandro and {Dekel}, Avishai and {Carollo}, C. Marcella and {Ceverino}, Daniel and {DeGraf}, Colin and {Lapiner}, Sharon and {Mandelker}, Nir and {Primack Joel}, R.},
        title = "{The confinement of star-forming galaxies into a main sequence through episodes of gas compaction, depletion and replenishment}",
      journal = {\mnras},
         year = 2016,
        month = apr,
       volume = {457},
       number = {3},
        pages = {2790-2813},
          doi = {10.1093/mnras/stw131},
archivePrefix = {arXiv},
       eprint = {1509.02529},
 primaryClass = {astro-ph.GA},
       adsurl = {https://ui.adsabs.harvard.edu/abs/2016MNRAS.457.2790T}
}

@article{Leys2013,
author = {{Leys}, Christophe and {Klein}, Olivier and {Bernard}, Philippe and {Licata}, Laurent},
title = {Detecting outliers: Do not use standard deviation around the mean, use absolute deviation around the median},
journal = {Journal of Experimental Social Psychology},
volume = {49},
number = {4},
pages = {764-766},
year = {2013},
issn = {0022-1031},
doi = {https://doi.org/10.1016/j.jesp.2013.03.013},
url = {https://www.sciencedirect.com/science/article/pii/S0022103113000668},
}

@ARTICLE{Birnboim2003,
       author = {{Birnboim}, Yuval and {Dekel}, Avishai},
        title = "{Virial shocks in galactic haloes?}",
      journal = {\mnras},
         year = 2003,
        month = oct,
       volume = {345},
       number = {1},
        pages = {349-364},
          doi = {10.1046/j.1365-8711.2003.06955.x},
archivePrefix = {arXiv},
       eprint = {astro-ph/0302161},
 primaryClass = {astro-ph},
       adsurl = {https://ui.adsabs.harvard.edu/abs/2003MNRAS.345..349B}
}

@ARTICLE{Romeo2023,
       author = {{Romeo}, Alessandro B. and {Agertz}, Oscar and {Renaud}, Florent},
        title = "{The specific angular momentum of disc galaxies and its connection with galaxy morphology, bar structure, and disc gravitational instability}",
      journal = {\mnras},
         year = 2023,
        month = jan,
       volume = {518},
       number = {1},
        pages = {1002-1021},
          doi = {10.1093/mnras/stac3074},
archivePrefix = {arXiv},
       eprint = {2204.02695},
 primaryClass = {astro-ph.GA},
       adsurl = {https://ui.adsabs.harvard.edu/abs/2023MNRAS.518.1002R}
}

@ARTICLE{Aubert2004,
       author = {{Aubert}, D. and {Pichon}, C. and {Colombi}, S.},
        title = "{The origin and implications of dark matter anisotropic cosmic infall on \raisebox{-0.5ex}\textasciitildeL$_{*}$ haloes}",
      journal = {\mnras},
         year = 2004,
        month = aug,
       volume = {352},
       number = {2},
        pages = {376-398},
          doi = {10.1111/j.1365-2966.2004.07883.x},
archivePrefix = {arXiv},
       eprint = {astro-ph/0402405},
 primaryClass = {astro-ph},
       adsurl = {https://ui.adsabs.harvard.edu/abs/2004MNRAS.352..376A}
}

@ARTICLE{Teyssier2011,
       author = {{Teyssier}, Romain and {Moore}, Ben and {Martizzi}, Davide and {Dubois}, Yohan and {Mayer}, Lucio},
        title = "{Mass distribution in galaxy clusters: the role of Active Galactic Nuclei feedback}",
      journal = {\mnras},
         year = 2011,
        month = jun,
       volume = {414},
       number = {1},
        pages = {195-208},
          doi = {10.1111/j.1365-2966.2011.18399.x},
archivePrefix = {arXiv},
       eprint = {1003.4744},
 primaryClass = {astro-ph.CO},
       adsurl = {https://ui.adsabs.harvard.edu/abs/2011MNRAS.414..195T}
}

@ARTICLE{Dubois2010,
       author = {{Dubois}, Yohan and {Devriendt}, Julien and {Slyz}, Adrianne and {Teyssier}, Romain},
        title = "{Jet-regulated cooling catastrophe}",
      journal = {\mnras},
         year = 2010,
        month = dec,
       volume = {409},
       number = {3},
        pages = {985-1001},
          doi = {10.1111/j.1365-2966.2010.17338.x},
archivePrefix = {arXiv},
       eprint = {1004.1851},
 primaryClass = {astro-ph.CO},
       adsurl = {https://ui.adsabs.harvard.edu/abs/2010MNRAS.409..985D}
}

@ARTICLE{Kimm2014,
       author = {{Kimm}, Taysun and {Cen}, Renyue},
        title = "{Escape Fraction of Ionizing Photons during Reionization: Effects due to Supernova Feedback and Runaway OB Stars}",
      journal = {\apj},
         year = 2014,
        month = jun,
       volume = {788},
       number = {2},
          eid = {121},
        pages = {121},
          doi = {10.1088/0004-637X/788/2/121},
archivePrefix = {arXiv},
       eprint = {1405.0552},
 primaryClass = {astro-ph.GA},
       adsurl = {https://ui.adsabs.harvard.edu/abs/2014ApJ...788..121K}
}

@ARTICLE{Dalgarno1972,
       author = {{Dalgarno}, A. and {McCray}, R.~A.},
        title = "{Heating and Ionization of HI Regions}",
      journal = {\araa},
         year = 1972,
        month = jan,
       volume = {10},
        pages = {375},
          doi = {10.1146/annurev.aa.10.090172.002111},
       adsurl = {https://ui.adsabs.harvard.edu/abs/1972ARA&A..10..375D}
}

@ARTICLE{Rosdahl2012,
       author = {{Rosdahl}, J. and {Blaizot}, J.},
        title = "{Extended Ly{\ensuremath{\alpha}} emission from cold accretion streams}",
      journal = {\mnras},
         year = 2012,
        month = jun,
       volume = {423},
       number = {1},
        pages = {344-366},
          doi = {10.1111/j.1365-2966.2012.20883.x},
archivePrefix = {arXiv},
       eprint = {1112.4408},
 primaryClass = {astro-ph.CO},
       adsurl = {https://ui.adsabs.harvard.edu/abs/2012MNRAS.423..344R}
}

@ARTICLE{Ciesla2023,
       author = {{Ciesla}, L. and {G{\'o}mez-Guijarro}, C. and {Buat}, V. and {Elbaz}, D. and {Jin}, S. and {B{\'e}thermin}, M. and {Daddi}, E. and {Franco}, M. and {Inami}, H. and {Magdis}, G. and {Magnelli}, B. and {Xiao}, M.},
        title = "{GOODS-ALMA 2.0: Last gigayear star formation histories of the so-called starbursts within the main sequence}",
      journal = {\aap},
         year = 2023,
        month = apr,
       volume = {672},
          eid = {A191},
        pages = {A191},
          doi = {10.1051/0004-6361/202245376},
archivePrefix = {arXiv},
       eprint = {2211.02510},
 primaryClass = {astro-ph.GA},
       adsurl = {https://ui.adsabs.harvard.edu/abs/2023A&A...672A.191C}
}

@ARTICLE{Kraljic2024,
       author = {{Kraljic}, Katarina and {Renaud}, Florent and {Dubois}, Yohan and {Pichon}, Christophe and {Agertz}, Oscar and {Andersson}, Eric and {Devriendt}, Julien and {Freundlich}, Jonathan and {Kaviraj}, Sugata and {Kimm}, Taysun and {Martin}, Garreth and {Peirani}, S{\'e}bastien and {Segovia Otero}, {\'A}lvaro and {Volonteri}, Marta and {Yi}, Sukyoung K.},
        title = "{Emergence and cosmic evolution of the Kennicutt-Schmidt relation driven by interstellar turbulence}",
      journal = {\aap},
         year = 2024,
        month = feb,
       volume = {682},
          eid = {A50},
        pages = {A50},
          doi = {10.1051/0004-6361/202347917},
archivePrefix = {arXiv},
       eprint = {2309.06485},
 primaryClass = {astro-ph.GA},
       adsurl = {https://ui.adsabs.harvard.edu/abs/2024A&A...682A..50K}
}

@ARTICLE{Renaud2022,
       author = {{Renaud}, Florent and {Segovia Otero}, {\'A}lvaro and {Agertz}, Oscar},
        title = "{The merger-starburst connection across cosmic times}",
      journal = {\mnras},
         year = 2022,
        month = nov,
       volume = {516},
       number = {4},
        pages = {4922-4931},
          doi = {10.1093/mnras/stac2557},
archivePrefix = {arXiv},
       eprint = {2209.03983},
 primaryClass = {astro-ph.GA},
       adsurl = {https://ui.adsabs.harvard.edu/abs/2022MNRAS.516.4922R}
}

@ARTICLE{Gomez2022,
       author = {{G{\'o}mez-Guijarro}, C. and {Elbaz}, D. and {Xiao}, M. and {Kokorev}, V.~I. and {Magdis}, G.~E. and {Magnelli}, B. and {Daddi}, E. and {Valentino}, F. and {Sargent}, M.~T. and {Dickinson}, M. and {B{\'e}thermin}, M. and {Franco}, M. and {Pope}, A. and {Kalita}, B.~S. and {Ciesla}, L. and {Demarco}, R. and {Inami}, H. and {Rujopakarn}, W. and {Shu}, X. and {Wang}, T. and {Zhou}, L. and {Alexander}, D.~M. and {Bournaud}, F. and {Chary}, R. and {Ferguson}, H.~C. and {Finkelstein}, S.~L. and {Giavalisco}, M. and {Iono}, D. and {Juneau}, S. and {Kartaltepe}, J.~S. and {Lagache}, G. and {Le Floc'h}, E. and {Leiton}, R. and {Leroy}, L. and {Lin}, L. and {Motohara}, K. and {Mullaney}, J. and {Okumura}, K. and {Pannella}, M. and {Papovich}, C. and {Treister}, E.},
        title = "{GOODS-ALMA 2.0: Starbursts in the main sequence reveal compact star formation regulating galaxy evolution prequenching}",
      journal = {\aap},
         year = 2022,
        month = mar,
       volume = {659},
          eid = {A196},
        pages = {A196},
          doi = {10.1051/0004-6361/202142352},
archivePrefix = {arXiv},
       eprint = {2201.02633},
 primaryClass = {astro-ph.GA},
       adsurl = {https://ui.adsabs.harvard.edu/abs/2022A&A...659A.196G}
}

@ARTICLE{Elbaz2007,
       author = {{Elbaz}, D. and {Daddi}, E. and {Le Borgne}, D. and {Dickinson}, M. and {Alexander}, D.~M. and {Chary}, R. -R. and {Starck}, J. -L. and {Brandt}, W.~N. and {Kitzbichler}, M. and {MacDonald}, E. and {Nonino}, M. and {Popesso}, P. and {Stern}, D. and {Vanzella}, E.},
        title = "{The reversal of the star formation-density relation in the distant universe}",
      journal = {\aap},
         year = 2007,
        month = jun,
       volume = {468},
       number = {1},
        pages = {33-48},
          doi = {10.1051/0004-6361:20077525},
archivePrefix = {arXiv},
       eprint = {astro-ph/0703653},
 primaryClass = {astro-ph},
       adsurl = {https://ui.adsabs.harvard.edu/abs/2007A&A...468...33E}
}

@ARTICLE{Moreno2021,
       author = {{Moreno}, Jorge and {Torrey}, Paul and {Ellison}, Sara L. and {Patton}, David R. and {Bottrell}, Connor and {Bluck}, Asa F.~L. and {Hani}, Maan H. and {Hayward}, Christopher C. and {Bullock}, James S. and {Hopkins}, Philip F. and {Hernquist}, Lars},
        title = "{Spatially resolved star formation and fuelling in galaxy interactions}",
      journal = {\mnras},
         year = 2021,
        month = may,
       volume = {503},
       number = {3},
        pages = {3113-3133},
          doi = {10.1093/mnras/staa2952},
archivePrefix = {arXiv},
       eprint = {2009.11289},
 primaryClass = {astro-ph.GA},
       adsurl = {https://ui.adsabs.harvard.edu/abs/2021MNRAS.503.3113M}
}

@ARTICLE{Segovia2022,
       author = {{Segovia Otero}, {\'A}lvaro and {Renaud}, Florent and {Agertz}, Oscar},
        title = "{VINTERGATAN IV: Cosmic phases of star formation in Milky Way-like galaxies}",
      journal = {\mnras},
         year = 2022,
        month = oct,
       volume = {516},
       number = {2},
        pages = {2272-2279},
          doi = {10.1093/mnras/stac2368},
       adsurl = {https://ui.adsabs.harvard.edu/abs/2022MNRAS.516.2272S}
}

@ARTICLE{Xu2021,
       author = {{Xu}, C.~K. and {Lisenfeld}, U. and {Gao}, Y. and {Renaud}, F.},
        title = "{NOEMA Observations of CO Emission in Arp 142 and Arp 238}",
      journal = {\apj},
         year = 2021,
        month = sep,
       volume = {918},
       number = {2},
          eid = {55},
        pages = {55},
          doi = {10.3847/1538-4357/ac0f77},
archivePrefix = {arXiv},
       eprint = {2106.15041},
 primaryClass = {astro-ph.GA},
       adsurl = {https://ui.adsabs.harvard.edu/abs/2021ApJ...918...55X}
}

@ARTICLE{Speagle2014,
       author = {{Speagle}, J.~S. and {Steinhardt}, C.~L. and {Capak}, P.~L. and {Silverman}, J.~D.},
        title = "{A Highly Consistent Framework for the Evolution of the Star-Forming ``Main Sequence'' from z \raisebox{-0.5ex}\textasciitilde 0-6}",
      journal = {\apjs},
         year = 2014,
        month = oct,
       volume = {214},
       number = {2},
          eid = {15},
        pages = {15},
          doi = {10.1088/0067-0049/214/2/15},
archivePrefix = {arXiv},
       eprint = {1405.2041},
 primaryClass = {astro-ph.GA},
       adsurl = {https://ui.adsabs.harvard.edu/abs/2014ApJS..214...15S}
}

@ARTICLE{Li2022,
       author = {{Li}, Hui and {Vogelsberger}, Mark and {Bryan}, Greg L. and {Marinacci}, Federico and {Sales}, Laura V. and {Torrey}, Paul},
        title = "{Formation and evolution of young massive clusters in galaxy mergers: the SMUGGLE view}",
      journal = {\mnras},
         year = 2022,
        month = jul,
       volume = {514},
       number = {1},
        pages = {265-279},
          doi = {10.1093/mnras/stac1136},
archivePrefix = {arXiv},
       eprint = {2109.10356},
 primaryClass = {astro-ph.GA},
       adsurl = {https://ui.adsabs.harvard.edu/abs/2022MNRAS.514..265L}
}

@ARTICLE{Elbaz2018,
       author = {{Elbaz}, D. and {Leiton}, R. and {Nagar}, N. and {Okumura}, K. and {Franco}, M. and {Schreiber}, C. and {Pannella}, M. and {Wang}, T. and {Dickinson}, M. and {D{\'\i}az-Santos}, T. and {Ciesla}, L. and {Daddi}, E. and {Bournaud}, F. and {Magdis}, G. and {Zhou}, L. and {Rujopakarn}, W.},
        title = "{Starbursts in and out of the star-formation main sequence}",
      journal = {\aap},
         year = 2018,
        month = aug,
       volume = {616},
          eid = {A110},
        pages = {A110},
          doi = {10.1051/0004-6361/201732370},
archivePrefix = {arXiv},
       eprint = {1711.10047},
 primaryClass = {astro-ph.GA},
       adsurl = {https://ui.adsabs.harvard.edu/abs/2018A&A...616A.110E}
}

@article{Muller2000,
author = {M{\"u}ller, J.W.},
year = {2000},
month = {07},
pages = {551},
title = {Possible Advantages of a Robust Evaluation of Comparisons},
volume = {105},
journal = {Journal of Research of the National Institute of Standards and Technology},
doi = {10.6028/jres.105.044}
}

@ARTICLE{Pillepich2019,
       author = {{Pillepich}, Annalisa and {Nelson}, Dylan and {Springel}, Volker and {Pakmor}, R{\"u}diger and {Torrey}, Paul and {Weinberger}, Rainer and {Vogelsberger}, Mark and {Marinacci}, Federico and {Genel}, Shy and {van der Wel}, Arjen and {Hernquist}, Lars},
        title = "{First results from the TNG50 simulation: the evolution of stellar and gaseous discs across cosmic time}",
      journal = {\mnras},
         year = 2019,
        month = dec,
       volume = {490},
       number = {3},
        pages = {3196-3233},
          doi = {10.1093/mnras/stz2338},
archivePrefix = {arXiv},
       eprint = {1902.05553},
 primaryClass = {astro-ph.GA},
       adsurl = {https://ui.adsabs.harvard.edu/abs/2019MNRAS.490.3196P}
}

@ARTICLE{Dubois2021,
       author = {{Dubois}, Yohan and {Beckmann}, Ricarda and {Bournaud}, Fr{\'e}d{\'e}ric and {Choi}, Hoseung and {Devriendt}, Julien and {Jackson}, Ryan and {Kaviraj}, Sugata and {Kimm}, Taysun and {Kraljic}, Katarina and {Laigle}, Clotilde and {Martin}, Garreth and {Park}, Min-Jung and {Peirani}, S{\'e}bastien and {Pichon}, Christophe and {Volonteri}, Marta and {Yi}, Sukyoung K.},
        title = "{Introducing the NEWHORIZON simulation: Galaxy properties with resolved internal dynamics across cosmic time}",
      journal = {\aap},
         year = 2021,
        month = jul,
       volume = {651},
          eid = {A109},
        pages = {A109},
          doi = {10.1051/0004-6361/202039429},
archivePrefix = {arXiv},
       eprint = {2009.10578},
 primaryClass = {astro-ph.GA},
       adsurl = {https://ui.adsabs.harvard.edu/abs/2021A&A...651A.109D}
}

@ARTICLE{Beraldo2020,
       author = {{Beraldo e Silva}, Leandro and {Debattista}, Victor P. and
         {Khachaturyants}, Tigran and {Nidever}, David},
        title = "{Geometric properties of galactic discs with clumpy episodes}",
      journal = {\mnras},
         year = 2020,
        month = mar,
       volume = {492},
       number = {4},
        pages = {4716-4726},
          doi = {10.1093/mnras/staa065},
archivePrefix = {arXiv},
       eprint = {1911.03717},
 primaryClass = {astro-ph.GA},
       adsurl = {https://ui.adsabs.harvard.edu/abs/2020MNRAS.492.4716B}
}

@ARTICLE{Renaud2019b,
       author = {{Renaud}, F. and {Bournaud}, F. and {Agertz}, O. and {Kraljic}, K. and
         {Schinnerer}, E. and {Bolatto}, A. and {Daddi}, E. and {Hughes}, A.},
        title = "{A diversity of starburst-triggering mechanisms in interacting galaxies and their signatures in CO emission}",
      journal = {\aap},
         year = "2019",
        month = "May",
       volume = {625},
          eid = {A65},
        pages = {A65},
          doi = {10.1051/0004-6361/201935222},
archivePrefix = {arXiv},
       eprint = {1902.02353},
 primaryClass = {astro-ph.GA},
       adsurl = {https://ui.adsabs.harvard.edu/abs/2019A&A...625A..65R}
}

@ARTICLE{Hopkins2013b,
       author = {{Hopkins}, Philip F. and {Cox}, Thomas J. and {Hernquist}, Lars and
         {Narayanan}, Desika and {Hayward}, Christopher C. and {Murray}, Norman},
        title = "{Star formation in galaxy mergers with realistic models of stellar feedback and the interstellar medium}",
      journal = {\mnras},
         year = "2013",
        month = "Apr",
       volume = {430},
        pages = {1901-1927},
          doi = {10.1093/mnras/stt017},
archivePrefix = {arXiv},
       eprint = {1206.0011},
 primaryClass = {astro-ph.CO},
       adsurl = {https://ui.adsabs.harvard.edu/\#abs/2013MNRAS.430.1901H}
}

@ARTICLE{Grisdale2019,
       author = {{Grisdale}, Kearn and {Agertz}, Oscar and {Renaud}, Florent and {Romeo}, Alessandro B. and {Devriendt}, Julien and {Slyz}, Adrianne},
        title = "{On the observed diversity of star formation efficiencies in Giant Molecular Clouds}",
      journal = {\mnras},
         year = 2019,
        month = jul,
       volume = {486},
       number = {4},
        pages = {5482-5491},
          doi = {10.1093/mnras/stz1201},
archivePrefix = {arXiv},
       eprint = {1902.00518},
 primaryClass = {astro-ph.GA},
       adsurl = {https://ui.adsabs.harvard.edu/abs/2019MNRAS.486.5482G}
}

@ARTICLE{Ohlin2019,
   author = {{Ohlin}, L. and {Renaud}, F. and {Agertz}, O.},
    title = "{Supernovae feedback propagation: the role of turbulence}",
  journal = {\mnras},
archivePrefix = "arXiv",
   eprint = {1902.00028},
     year = 2019,
    month = may,
   volume = 485,
    pages = {3887-3894},
      doi = {10.1093/mnras/stz705},
   adsurl = {http://cdsads.u-strasbg.fr/abs/2019MNRAS.485.3887O}
}

@ARTICLE{Renaud2019,
   author = {{Renaud}, F. and {Bournaud}, F. and {Daddi}, E. and {Wei{\ss}}, A.
  },
    title = "{Three regimes of CO emission in galaxy mergers}",
  journal = {\aap},
archivePrefix = "arXiv",
   eprint = {1811.06547},
     year = 2019,
    month = jan,
   volume = 621,
      eid = {A104},
    pages = {A104},
      doi = {10.1051/0004-6361/201834397},
   adsurl = {http://adsabs.harvard.edu/abs/2019A%26A...621A.104R}
}

@Article{agertz2015,
  author	= {{Agertz}, O. and {Kravtsov}, A.~V.},
  title		= "{On the Interplay between Star Formation and Feedback in
		  Galaxy Formation Simulations}",
  journal	= {\apj},
  archiveprefix	= "arXiv",
  eprint	= {1404.2613},
  year		= 2015,
  month		= may,
  volume	= 804,
  eid		= {18},
  pages		= {18},
  doi		= {10.1088/0004-637X/804/1/18},
  adsurl	= {http://adsabs.harvard.edu/abs/2015ApJ...804...18A}
}

@Article{cortijo2017,
  author	= {{Cortijo-Ferrero}, C. and {Gonz{\'a}lez Delgado}, R.~M.
		  and {P{\'e}rez}, E. and {Cid Fernandes}, R. and
		  {S{\'a}nchez}, S.~F. and {de Amorim}, A.~L. and {Di
		  Matteo}, P. and {Garc{\'{\i}}a-Benito}, R. and {Lacerda},
		  E.~A.~D. and {L{\'o}pez Fern{\'a}ndez}, R. and {Tadhunter}, C. },
  title		= "{Star formation histories in mergers: The spatially
		  resolved properties of the early-stage merger LIRGs IC 1623
		  and NGC 6090}",
  journal	= {ArXiv e-prints},
  archiveprefix	= "arXiv",
  eprint	= {1702.06544},
  year		= 2017,
  month		= feb,
  adsurl	= {http://adsabs.harvard.edu/abs/2017arXiv170206544C}
}

@Article{cox2008,
  author	= {{Cox}, T.~J. and {Jonsson}, P. and {Somerville}, R.~S. and
		  {Primack}, J.~R. and {Dekel}, A.},
  title		= "{The effect of galaxy mass ratio on merger-driven
		  starbursts}",
  journal	= {\mnras},
  archiveprefix	= "arXiv",
  eprint	= {0709.3511},
  year		= 2008,
  month		= feb,
  volume	= 384,
  pages		= {386-409},
  doi		= {10.1111/j.1365-2966.2007.12730.x},
  adsurl	= {http://adsabs.harvard.edu/abs/2008MNRAS.384..386C}
}

@Article{dekel2006,
  author	= {{Dekel}, A. and {Birnboim}, Y.},
  title		= "{Galaxy bimodality due to cold flows and shock heating}",
  journal	= {\mnras},
  eprint	= {arXiv:astro-ph/0412300},
  year		= 2006,
  month		= may,
  volume	= 368,
  pages		= {2-20},
  doi		= {10.1111/j.1365-2966.2006.10145.x},
  adsurl	= {http://adsabs.harvard.edu/abs/2006MNRAS.368....2D}
}

@Article{dimatteo2008,
  author	= {{Di Matteo}, P. and {Bournaud}, F. and {Martig}, M. and
		  {Combes}, F. and {Melchior}, A.-L. and {Semelin}, B.},
  title		= "{On the frequency, intensity, and duration of starburst
		  episodes triggered by galaxy interactions and mergers}",
  journal	= {\aap},
  archiveprefix	= "arXiv",
  eprint	= {0809.2592},
  year		= 2008,
  month		= dec,
  volume	= 492,
  pages		= {31-49},
  doi		= {10.1051/0004-6361:200809480},
  adsurl	= {http://adsabs.harvard.edu/abs/2008A%26A...492...31D}
}

@Article{ellison2008,
  author	= {{Ellison}, S.~L. and {Patton}, D.~R. and {Simard}, L. and
		  {McConnachie}, A.~W. },
  title		= "{Galaxy Pairs in the Sloan Digital Sky Survey. I. Star
		  Formation, Active Galactic Nucleus Fraction, and the
		  Mass-Metallicity Relation}",
  journal	= {\aj},
  archiveprefix	= "arXiv",
  eprint	= {0803.0161},
  year		= 2008,
  month		= may,
  volume	= 135,
  pages		= {1877-1899},
  doi		= {10.1088/0004-6256/135/5/1877},
  adsurl	= {http://adsabs.harvard.edu/abs/2008AJ....135.1877E}
}

@Article{federrath2012,
  author	= {{Federrath}, C. and {Klessen}, R.~S.},
  title		= "{The Star Formation Rate of Turbulent Magnetized Clouds:
		  Comparing Theory, Simulations, and Observations}",
  journal	= {\apj},
  archiveprefix	= "arXiv",
  eprint	= {1209.2856},
  primaryclass	= "astro-ph.SR",
  year		= 2012,
  month		= dec,
  volume	= 761,
  eid		= {156},
  pages		= {156},
  doi		= {10.1088/0004-637X/761/2/156},
  adsurl	= {http://adsabs.harvard.edu/abs/2012ApJ...761..156F}
}

@Article{fensch2017,
  author	= {{Fensch}, J. and {Renaud}, F. and {Bournaud}, F. and
		  {Duc}, P.-A. and {Agertz}, O. and {Amram}, P. and {Combes},
		  F. and {Di Matteo}, P. and {et al.}},
  title		= "{High-redshift major mergers weakly enhance star
		  formation}",
  journal	= {\mnras},
  archiveprefix	= "arXiv",
  eprint	= {1610.03877},
  year		= 2017,
  month		= feb,
  volume	= 465,
  pages		= {1934-1949},
  doi		= {10.1093/mnras/stw2920},
  adsurl	= {http://adsabs.harvard.edu/abs/2017MNRAS.465.1934F}
}

@Article{grisdale2017,
  author	= {{Grisdale}, K. and {Agertz}, O. and {Romeo}, A.~B. and
		  {Renaud}, F. and {Read}, J.~I.},
  title		= "{The impact of stellar feedback on the density and
		  velocity structure of the interstellar medium}",
  journal	= {\mnras},
  archiveprefix	= "arXiv",
  eprint	= {1608.08639},
  year		= 2017,
  month		= apr,
  volume	= 466,
  pages		= {1093-1110},
  doi		= {10.1093/mnras/stw3133},
  adsurl	= {http://adsabs.harvard.edu/abs/2017MNRAS.466.1093G}
}

@Article{haardt1996,
  author	= {{Haardt}, F. and {Madau}, P.},
  title		= "{Radiative Transfer in a Clumpy Universe. II. The
		  Ultraviolet Extragalactic Background}",
  journal	= {\apj},
  eprint	= {arXiv:astro-ph/9509093},
  year		= 1996,
  month		= apr,
  volume	= 461,
  pages		= {20},
  doi		= {10.1086/177035},
  adsurl	= {http://adsabs.harvard.edu/abs/1996ApJ...461...20H}
}

@Article{hennebelle2011,
  author	= {{Hennebelle}, P. and {Chabrier}, G.},
  title		= "{Analytical Star Formation Rate from Gravoturbulent
		  Fragmentation}",
  journal	= {\apjl},
  archiveprefix	= "arXiv",
  eprint	= {1110.0033},
  primaryclass	= "astro-ph.GA",
  year		= 2011,
  month		= dec,
  volume	= 743,
  eid		= {L29},
  pages		= {L29},
  doi		= {10.1088/2041-8205/743/2/L29},
  adsurl	= {http://adsabs.harvard.edu/abs/2011ApJ...743L..29H}
}

@Article{karl2010,
  author	= {{Karl}, S.~J. and {Naab}, T. and {Johansson}, P.~H. and
		  {Kotarba}, H. and {Boily}, C.~M. and {Renaud}, F. and
		  {Theis}, C.},
  title		= "{One Moment in Time Modeling Star Formation in the
		  Antennae}",
  journal	= {\apjl},
  archiveprefix	= "arXiv",
  eprint	= {1003.0685},
  primaryclass	= "astro-ph.CO",
  year		= 2010,
  month		= jun,
  volume	= 715,
  pages		= {L88-L93},
  doi		= {10.1088/2041-8205/715/2/L88},
  adsurl	= {http://adsabs.harvard.edu/abs/2010ApJ...715L..88K}
}

@Article{krumholz2005,
  author	= {{Krumholz}, M.~R. and {McKee}, C.~F.},
  title		= "{A General Theory of Turbulence-regulated Star Formation,
		  from Spirals to Ultraluminous Infrared Galaxies}",
  journal	= {\apj},
  eprint	= {arXiv:astro-ph/0505177},
  year		= 2005,
  month		= sep,
  volume	= 630,
  pages		= {250-268},
  doi		= {10.1086/431734},
  adsurl	= {http://adsabs.harvard.edu/abs/2005ApJ...630..250K}
}

@Article{lee2016,
  author	= {{Lee}, Y.-N. and {Hennebelle}, P.},
  title		= "{Formation of a protocluster: A virialized structure from
		  gravoturbulent collapse. II. A two-dimensional analytical
		  model for a rotating and accreting system}",
  journal	= {\aap},
  archiveprefix	= "arXiv",
  eprint	= {1603.07983},
  primaryclass	= "astro-ph.SR",
  year		= 2016,
  month		= jun,
  volume	= 591,
  eid		= {A31},
  pages		= {A31},
  doi		= {10.1051/0004-6361/201527982},
  adsurl	= {http://adsabs.harvard.edu/abs/2016A%26A...591A..31L}
}

@article{Lotz2011,
  author        = {{Lotz}, Jennifer M. and {Jonsson}, Patrik and {Cox}, T.~J. and {Croton}, Darren and {Primack}, Joel R. and {Somerville}, Rachel S. and {Stewart}, Kyle},
  title         = {{The Major and Minor Galaxy Merger Rates at z < 1.5}},
  journal       = {\apj},
  year          = 2011,
  month         = dec,
  volume        = {742},
  number        = {2},
  eid           = {103},
  pages         = {103},
  doi           = {10.1088/0004-637X/742/2/103},
  archiveprefix = {arXiv},
  eprint        = {1108.2508},
  primaryclass  = {astro-ph.CO},
  adsurl        = {https://ui.adsabs.harvard.edu/abs/2011ApJ...742..103L}
}

@Article{mihos1996,
  author	= {{Mihos}, J.~C. and {Hernquist}, L.},
  title		= "{Gasdynamics and Starbursts in Major Mergers}",
  journal	= {\apj},
  eprint	= {arXiv:astro-ph/9512099},
  year		= 1996,
  month		= jun,
  volume	= 464,
  pages		= {641-+},
  doi		= {10.1086/177353},
  adsurl	= {http://adsabs.harvard.edu/abs/1996ApJ...464..641M}
}

@Article{mineo2014,
  author	= {{Mineo}, S. and {Rappaport}, S. and {Levine}, A. and
		  {Pooley}, D. and {Steinhorn}, B. and {Homan}, J.},
  title		= "{A Comprehensive X-Ray and Multiwavelength Study of the
		  Colliding Galaxy Pair NGC 2207/IC 2163}",
  journal	= {\apj},
  archiveprefix	= "arXiv",
  eprint	= {1410.2472},
  year		= 2014,
  month		= dec,
  volume	= 797,
  eid		= {91},
  pages		= {91},
  doi		= {10.1088/0004-637X/797/2/91},
  adsurl	= {http://adsabs.harvard.edu/abs/2014ApJ...797...91M}
}

@Article{naab2017,
  author	= {{Naab}, T. and {Ostriker}, J.~P.},
  title		= "{Theoretical Challenges in Galaxy Formation}",
  journal	= {\araa},
  archiveprefix	= "arXiv",
  eprint	= {1612.06891},
  year		= 2017,
  month		= aug,
  volume	= 55,
  pages		= {59-109},
  doi		= {10.1146/annurev-astro-081913-040019},
  adsurl	= {http://adsabs.harvard.edu/abs/2017ARA%26A..55...59N}
}

@Article{noeske2007,
  author        = {{Noeske}, K.~G. and {Weiner}, B.~J. and {Faber}, S.~M. and {Papovich}, C. and {Koo}, D.~C. and {Somerville}, R.~S. and {Bundy}, K. and {Conselice}, C.~J. and {Newman}, J.~A. and {Schiminovich}, D. and {Le Floc'h}, E. and {Coil}, A.~L. and {Rieke}, G.~H. and {Lotz}, J.~M. and {Primack}, J.~R. and {Barmby}, P. and {Cooper}, M.~C. and {Davis}, M. and {Ellis}, R.~S. and {Fazio}, G.~G. and {Guhathakurta}, P. and {Huang}, J. and {Kassin}, S.~A. and {Martin}, D.~C. and {Phillips}, A.~C. and {Rich}, R.~M. and {Small}, T.~A. and {Willmer}, C.~N.~A. and {Wilson}, G.},
  title         = {{Star Formation in AEGIS Field Galaxies since z=1.1: The Dominance of Gradually Declining Star Formation, and the Main Sequence of Star-forming Galaxies}},
  journal       = {\apjl},
  year          = 2007,
  month         = may,
  volume        = {660},
  number        = {1},
  pages         = {L43-L46},
  doi           = {10.1086/517926},
  archiveprefix = {arXiv},
  eprint        = {astro-ph/0701924},
  primaryclass  = {astro-ph},
  adsurl        = {https://ui.adsabs.harvard.edu/abs/2007ApJ...660L..43N}
}

@Article{padoan2011,
  author	= {{Padoan}, P. and {Nordlund}, {\AA}.},
  title		= "{The Star Formation Rate of Supersonic Magnetohydrodynamic
		  Turbulence}",
  journal	= {\apj},
  archiveprefix	= "arXiv",
  eprint	= {0907.0248},
  primaryclass	= "astro-ph.GA",
  year		= 2011,
  month		= mar,
  volume	= 730,
  eid		= {40},
  pages		= {40},
  doi		= {10.1088/0004-637X/730/1/40},
  adsurl	= {http://adsabs.harvard.edu/abs/2011ApJ...730...40P}
}

@Article{perret2014,
  author	= {{Perret}, V. and {Renaud}, F. and {Epinat}, B. and
		  {Amram}, P. and {Bournaud}, F. and {Contini}, T. and
		  {Teyssier}, R. and {Lambert}, J.-C.},
  title		= "{Evolution of the mass, size, and star formation rate in
		  high redshift merging galaxies. MIRAGE - A new sample of
		  simulations with detailed stellar feedback}",
  journal	= {\aap},
  archiveprefix	= "arXiv",
  eprint	= {1307.7130},
  primaryclass	= "astro-ph.CO",
  year		= 2014,
  month		= feb,
  volume	= 562,
  eid		= {A1},
  pages		= {A1},
  doi		= {10.1051/0004-6361/201322395},
  adsurl	= {http://adsabs.harvard.edu/abs/2014A%26A...562A...1P}
}

@Article{renaud2013b,
  author	= {{Renaud}, F. and {Bournaud}, F. and {Emsellem}, E. and
		  {Elmegreen}, B. and {Teyssier}, R. and {Alves}, J. and
		  {Chapon}, D. and {Combes}, F. and {et al.}},
  title		= "{A sub-parsec resolution simulation of the Milky Way:
		  global structure of the interstellar medium and properties
		  of molecular clouds}",
  journal	= {\mnras},
  archiveprefix	= "arXiv",
  eprint	= {1307.5639},
  primaryclass	= "astro-ph.GA",
  year		= 2013,
  month		= dec,
  volume	= 436,
  pages		= {1836-1851},
  doi		= {10.1093/mnras/stt1698},
  adsurl	= {http://adsabs.harvard.edu/abs/2013MNRAS.436.1836R}
}

@Article{renaud2014b,
  author	= {{Renaud}, F. and {Bournaud}, F. and {Kraljic}, K. and
		  {Duc}, P.-A.},
  title		= "{Starbursts triggered by intergalactic tides
		  andinterstellar compressive turbulence}",
  journal	= {\mnras},
  archiveprefix	= "arXiv",
  eprint	= {1403.7316},
  year		= 2014,
  month		= jul,
  volume	= 442,
  pages		= {L33-L37},
  doi		= {10.1093/mnrasl/slu050},
  adsurl	= {http://adsabs.harvard.edu/abs/2014MNRAS.442L..33R}
}

@Article{rodighiero2011,
  author	= {{Rodighiero}, G. and {Daddi}, E. and {Baronchelli}, I. and
		  {Cimatti}, A. and {Renzini}, A. and {Aussel}, H. and
		  {Popesso}, P. and {Lutz}, D. and {et al.}},
  title		= "{The Lesser Role of Starbursts in Star Formation at z =
		  2}",
  journal	= {\apjl},
  archiveprefix	= "arXiv",
  eprint	= {1108.0933},
  primaryclass	= "astro-ph.CO",
  year		= 2011,
  month		= oct,
  volume	= 739,
  eid		= {L40},
  pages		= {L40},
  doi		= {10.1088/2041-8205/739/2/L40},
  adsurl	= {http://adsabs.harvard.edu/abs/2011ApJ...739L..40R}
}

@Article{schreiber2015,
  author	= {{Schreiber}, C. and {Pannella}, M. and {Elbaz}, D. and
		  {B{\'e}thermin}, M. and {Inami}, H. and {Dickinson}, M. and
		  {Magnelli}, B. and {Wang}, T. and {et al.}},
  title		= "{The Herschel view of the dominant mode of galaxy growth
		  from z = 4 to the present day}",
  journal	= {\aap},
  archiveprefix	= "arXiv",
  eprint	= {1409.5433},
  year		= 2015,
  month		= mar,
  volume	= 575,
  eid		= {A74},
  pages		= {A74},
  doi		= {10.1051/0004-6361/201425017},
  adsurl	= {http://adsabs.harvard.edu/abs/2015A%26A...575A..74S}
}

@InProceedings{schweizer2005,
  author	= {{Schweizer}, F.},
  title		= "{Merger-Induced Starbursts}",
  booktitle	= {ASSL Vol. 329: Starbursts: From 30 Doradus to Lyman Break
		  Galaxies},
  year		= 2005,
  month		= may,
  pages		= {143},
  adsurl	= {http://adsabs.harvard.edu/cgi-bin/nph-bib_query?bibcode=2005sdlb.proc..143S&db_key=AST}
}

@Article{sutherland1993,
  author	= {{Sutherland}, R.~S. and {Dopita}, M.~A.},
  title		= "{Cooling functions for low-density astrophysical plasmas}",
  journal	= {\apjs},
  year		= 1993,
  month		= sep,
  volume	= 88,
  pages		= {253-327},
  doi		= {10.1086/191823},
  adsurl	= {http://adsabs.harvard.edu/abs/1993ApJS...88..253S}
}

@Article{teyssier2002,
  author	= {{Teyssier}, R.},
  title		= "{Cosmological hydrodynamics with adaptive mesh refinement.
		  A new high resolution code called RAMSES}",
  journal	= {\aap},
  eprint	= {arXiv:astro-ph/0111367},
  year		= 2002,
  month		= apr,
  volume	= 385,
  pages		= {337-364},
  doi		= {10.1051/0004-6361:20011817},
  adsurl	= {http://adsabs.harvard.edu/abs/2002A%26A...385..337T}
}

@Article{teyssier2010,
  author	= {{Teyssier}, R. and {Chapon}, D. and {Bournaud}, F.},
  title		= "{The Driving Mechanism of Starbursts in Galaxy Mergers}",
  journal	= {\apjl},
  archiveprefix	= "arXiv",
  eprint	= {1006.4757},
  primaryclass	= "astro-ph.CO",
  year		= 2010,
  month		= sep,
  volume	= 720,
  pages		= {L149-L154},
  doi		= {10.1088/2041-8205/720/2/L149},
  adsurl	= {http://adsabs.harvard.edu/abs/2010ApJ...720L.149T}
}

@Article{whitmore2010,
  author	= {{Whitmore}, B.~C. and {Chandar}, R. and {Schweizer}, F.
		  and {Rothberg}, B. and {Leitherer}, C. and {Rieke}, M. and
		  {Rieke}, G. and {Blair}, W.~P. and {Mengel}, S. and
		  {Alonso-Herrero}, A.},
  title		= "{The Antennae Galaxies (NGC 4038/4039) Revisited: Advanced
		  Camera for Surveys and NICMOS Observations of a
		  Prototypical Merger}",
  journal	= {\aj},
  archiveprefix	= "arXiv",
  eprint	= {1005.0629},
  primaryclass	= "astro-ph.EP",
  year		= 2010,
  month		= jul,
  volume	= 140,
  pages		= {75-109},
  doi		= {10.1088/0004-6256/140/1/75},
  adsurl	= {http://adsabs.harvard.edu/abs/2010AJ....140...75W}
}

@PREAMBLE{ {\providecommand{\noopsort}[1]{}} }

\appendix
\section{Examples of evolutionary tracks of SBMS galaxies}
\label{sec:examples}

\begin{figure*}
\centering
\includegraphics{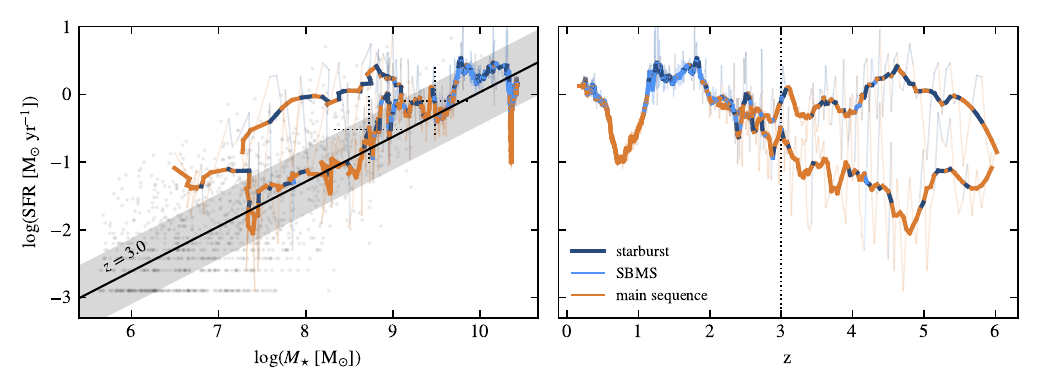}
\caption{\emph{Left panel:} Evolution of two examples of SBMS galaxies identified as such at $z=3$ in the stellar mass-SFR plane. The positions of the galaxies at $z=3$ are indicated by the large plus signs. \emph{Right panel:} Evolution of the SFR as function of redshift of the same galaxies. The curves are color-coded by the category of the galaxy: SBMS in light blue, off-main sequence starburst in dark blue, and main sequence in orange. The curves have been smoothed by a Savitzky-Golay algorithm to improve readability, with the original measurements shown by the semi-transparent lines. Black dots represent the entire population of star forming galaxies at $z=3$, with the main sequence shown with the solid line and shaded area. We remind the reader that the position of the main sequence changes with redshift: these elements in black can be used to guide the eye only at $z=3$.}
\label{fig:track}
\end{figure*}

Examining the evolution of every galaxy is beyond the scope of this paper. However, to illustrate possible paths and variations along them, \fig{track} shows the history of two galaxies detected as SBMS at $z=3$. The case with the lowest SFR disappears at $z\approx 2.3$ when it merges with a larger galaxy. The rapid variations of the SFR visible in these examples (and in most, if not all simulated SFRs) explain the short durations of SBMS episodes.

\section{Merger frequency}

\begin{figure}
\centering
\includegraphics{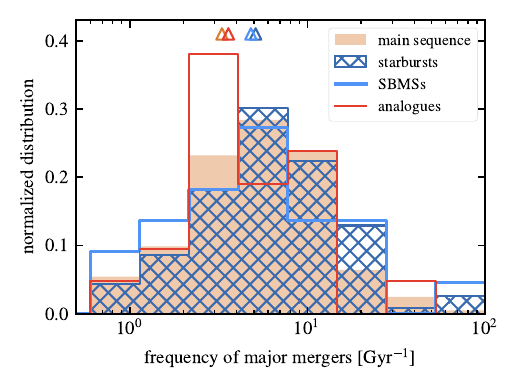}
\caption{Frequency of major mergers computed as the number of major mergers divided by the lifetime of the galaxy (at $z=3$). Triangles indicate the median value of each class.}
\label{fig:MMfrequency}
\end{figure}

\fig{MMfrequency} shows that starburst and SBMS galaxies experience major mergers $\approx 1.4$ time more frequently than the other galaxies. The more numerous and more frequent major merger events favor an enhanced star formation activity, and a rapid build-up of the stellar component (by accretion and accelerated in situ formation).

With the mass growth, this aspect is one of the main variation with redshift, caused by the evolution of the large-scale environment. The analysis is repeated at $z=1$ in \fig{MMfrequency_z1}, and completed with the count of merger events in \fig{nmergers_z1}. As detailed in \sect{redshift}, the peculiarity of SBMS galaxies in terms of merger events fade with decreasing redshift, further highlighting the early and precocious nature of their assembly.

\begin{figure}
\centering
\includegraphics{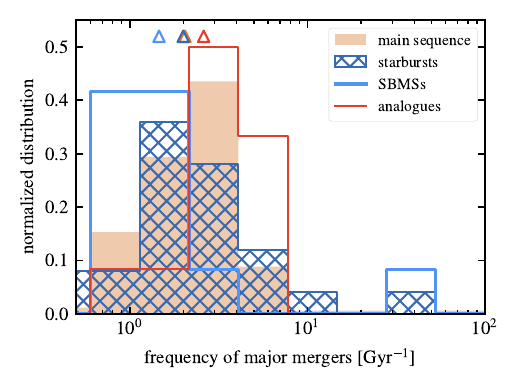}
\caption{Same as \fig{MMfrequency} but at $z=1$.}
\label{fig:MMfrequency_z1}
\end{figure}

\begin{figure}
\centering
\includegraphics{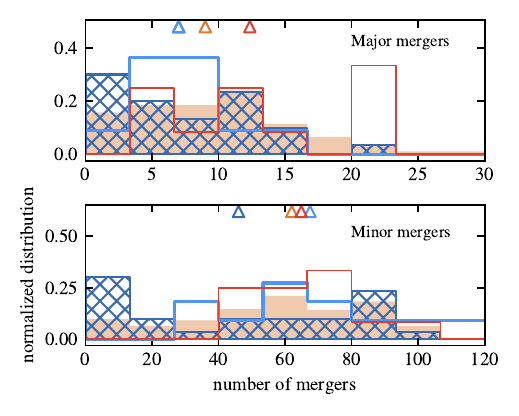}
\caption{Same as \fig{nmergers} but at $z=1$.}
\label{fig:nmergers_z1}
\end{figure}

\end{document}